\documentclass[useAMS,usenatbib]{mn2e}

\usepackage{graphicx} 
\usepackage{caption}
\usepackage{tabularx}
\usepackage{xcolor}
\usepackage{soul}
\newcommand{\change}[1]{#1}
\def\note #1]{{\bf #1]}}
\def\note #1]{{\bf #1]}}

\newcommand{\nuMax}{$\nu_{\rm max}$}
\newcommand{\DeltaNu}{$\Delta\nu$}
\newcommand{\eps}{$\epsilon$}

\newcommand{\PL}{75\,s} 

\newcommand{\uHz}[1]{$#1\,\mu \rm Hz$}

\newcommand{\all}{6661} 
\newcommand{\DPi}{\Delta\Pi_1}
\newcommand{\Ms}{$M_{\odot}$}
\newcommand{\logg}{log(\textit{g})}
\newcommand{\Teff}{$\rm T_{eff}$}
\newcommand{\rc}{red clump}
\newcommand{\rgb}{red-giant branch}
\newcommand{\scl}{secondary clump}
\newcommand{\agb}{asymptotic giant branch}
\newcommand{\loggLimit}{2.38}
\newcommand{\Kusz}{Kuszlewicz et al. (in prep.)}

\title[Evolutionary State of Red-Giant Stars]{Insights from the APOKASC Determination of the Evolutionary State of Red-Giant Stars by consolidation of different methods}
\author[Elsworth et al. ]{
Yvonne Elsworth$^{1,2}$\thanks{E-mail:y.p.elsworth@bham.ac.uk},
Saskia Hekker$^{3,2}$,
Jennifer A. Johnson$^{4},$
Thomas Kallinger$^{5}$,\newauthor
Benoit Mosser$^{6}$,
Marc Pinsonneault$^{4}$,
Marc Hon$^{7}$,
James Kuszlewicz$^{3,2}$,
Andrea Miglio$^{1,2}$,\newauthor
Aldo Serenelli$^{8,9}$,
Dennis Stello$^{2,7,10}$,
Jamie Tayar$^{11,12}$,
Mathieu Vrard$^{13}$\\
$^{1}$School of Physics and Astronomy, University of Birmingham, Birmingham B15 2TT, UK\\
$^{2}$Stellar Astrophysics Centre, Department of Physics and Astronomy, Aarhus University, Ny Munkegade 120, DK-8000 Aarhus C,
Denmark\\
$^{3}$Max-Planck-Institut f{\"u}r Sonnensystemforschung, Justus-von-Liebig-Weg 3, 37077 G{\"o}ttingen, Germany\\
$^{4}$Department of Astronomy, The Ohio State University, Columbus, OH 43210, USA \\
$^{5}$Institut f\"ur Astrophysik, Universit\"at Wien, T\"urkenschanzstrasse 17, 1180 Vienna, Austria\\
$^{6}$LESIA, Observatoire de Paris, Universit\'e PSL, CNRS, Sorbonne Universit\'e, Universit\'e de Paris, 5 place Jules Janssen, 92195 Meudon, France \\
$^{7}$School of Physics, University of New South Wales, NSW 2052, Australia\\
$^{8}$Institute of Space Sciences (ICE, CSIC) Campus UAB, Carrer de Can Magrans, s/n, E-08193, Bellaterra, Spain \\
$^{9}$Institut d'Estudis Espacials de Catalunya (IEEC), C/Gran Capita, 2-4, E-08034, Barcelona, Spain\\
$^{10}$Sydney Institute for Astronomy (SIfA), School of Physics, University of Sydney, NSW 2006, Australia\\
$^{11}$Institute for Astronomy, University of Hawaii, 2680 Woodlawn Drive, Honolulu, Hawaii 96822, USA\\
$^{12}$Hubble Fellow\\
$^{13}$Instituto de Astrof\'{\i}sica e Ci\^{e}ncias do Espa\c{c}o, Universidade do Porto, CAUP, Rua das Estrelas, 4150-762 Porto, Portugal\\
}
\begin{document}


\maketitle

\label{firstpage}

\begin{abstract}
The internal working of low-mass stars is of great significance to both the study of stellar structure and the history of the Milky Way.
Asteroseismology has the power to directly sense the internal structure of stars and allows for the determination of the evolutionary state -- i.e. has helium burning commenced or is the energy generated only by the fusion in the hydrogen-burning shell?
We use  observational data  from red-giant stars in 
a combination (known as APOKASC) of asteroseismology (from the \textit{Kepler} mission) and spectroscopy (from SDSS/APOGEE). 
The new feature of the analysis is that the APOKASC evolutionary state determination is based on the comparison of diverse approaches to the investigation  of the frequency-power spectrum.
The high level of agreement between the methods is a strong validation of the approaches.
Stars for which there is not a consensus view are readily identified. 
The comparison also facilitates the identification of unusual stars including those that show evidence for very strong coupling between p and g cavities.
The comparison between the  classification based on the spectroscopic data and asteroseismic data have led to a new value for the statistical uncertainty in APOGEE temperatures.
These consensus evolutionary states will be used as an input for methods that derive masses and ages for these stars based on comparison of observables with stellar evolutionary models (`grid-based modeling') and as a training set for machine-learning and other data-driven methods of evolutionary state determination. 
\end{abstract}

\section{Introduction}

Asteroseismology coupled with spectroscopic analysis of the light from red-giant stars is a
very powerful tool for probing the structure of the Galaxy, and  for
testing and improving models of stellar structure and evolution.
The effectiveness of this approach  has been widely discussed in the literature (for example, see \cite{hekkerJCD2017} and references therein).

Informative  diagnostics of stellar populations and stellar physics are available
if core-helium-burning stars (CHeB) can reliably be separated from first
ascent red giant (RGB) stars in the field.
However this is not always easy.
Most CHeB stars have similar visual absolute magnitudes and
form features in the HR diagram known as the \rc\ (RC) or as the horizontal branch (HB).
If the observed stars have a range of distances, ages, and metallicities, the locations of CHeB and RGB stars can overlap in the color-apparent magnitude diagram.
The resulting difficulty in disentangling the populations deprives
us of crucial insights.


A vitally important solution to this problem is provided by asteroseismology.
In many cases asteroseismology is able to distinguish the evolutionary
states (see references given later in this section) -- with and
without core-helium-burning -- and it opens up the opportunity for much more
precision in these important areas of research. Several techniques based
on understanding the oscillations in stars with different structure and the
effect on the frequency power spectrum have been developed. Each technique
has its strengths and limitations. In the past, these methods have been
considered in isolation. This paper presents the first time that multiple
methods are used to determine the evolutionary state of any given star and
hence the restrictions in any given method are mitigated. The evolutionary
states reported here are used to train machine-learning methods that will
be increasingly important as the number of lightcurves for red giants grows to over $10^5$.

We start this paper by giving some examples of current projects where clear separation of RGB and CHeB stars is important.
For example, mass loss on the upper giant branch can be inferred if pre- and post-helium-flash stars
can be identified \citep{miglio2012}.
\citet{ibenRood1969} used the fact that  the lifetime of stars on the \rgb\ above the red-giant bump is dependent on the initial helium abundance and  the lifetime of the \rc\ or HB phase is relatively insensitive to it to  show that the ratio of the number of \rgb\ stars above the magnitude of the \rc\ to the
number of \rc\ and HB  stars is a measure of the helium abundance of the population.
This is known as the R-method.
It is used extensively
in globular clusters \change{\citep{buzzoni1983,constantino2015}}, whose cool red giants do not feature reliable helium spectral features, and  has also been applied in the bulge \citep{renzini1994, minniti1995, tiede1995}.

Both cluster and bulge work relies on the distance to all the stars being essentially the same and
the age and/or metallicity range of the population being narrow so that the CHeB and the
RGB stars can be identified by their position on the color-magnitude diagram.
A technique that does not depend solely on temperature and apparent magnitude to identify evolutionary states is therefore desirable.

During evolution on the red-giant branch, stars pass through a stage with a very extensive convective envelope. 
During this stage, material that has had its composition changed by fusion is mixed with unprocessed surface material.
The quantification of the amount of mixing is another property that can only
be measured if stars that have passed through the entire first ascent red giant
branch can be separated from stars that are still working their way up the
giant branch. For example, \citet{masseron2017a} used the evolutionary states
of \cite{elsworth2017evolYE} and the abundances from \cite{hawkins2016} to argue that
unexpected dredge-up of newly produced carbon near the tip of the RGB was a possible
explanation of lower [C/N] ratios in the RC than stars with similar metallicity at
the tip of the RGB.

The nature of companions around CHeB stars will be different from those around RGB stars.
CHeB stars have passed through a phase  of very inflated radii (by comparison with lower giant-branch RGB stars).
Hence,  close-in companions, including planets, are likely to have been swallowed more frequently for CHeB stars than for RGB stars.
An engulfment will also affect the subsequent rotation of the star, as will any transfer
of angular momentum inside the star on the upper giant branch and the shrinking
of the envelope after the ignition of helium.

Identification of \scl\ stars \citep{girardi1999}, which ignite core-helium burning
in non-degenerate conditions before reaching the tip of the \rgb\ and therefore
have lower luminosities on average than the RC, can be used to isolate a population of
relatively young stars.
\citet{casagrande2016} noted the decreasing fraction of \scl\ stars with height above the plane, confirming a vertical age gradient in the Milky Way disk.

Additionally, there is the surprising, and still poorly understood, effect that there is an evolutionary-state dependent offset \citep{pinsonneault2014,pinsonneault2018} between asteroseismic and spectroscopic surface gravities  for \logg $>$ \loggLimit.
That means that there is the need for a correction to the spectroscopically determined \logg\ the value of which depends on whether the star is on the \rgb\ or in the \rc.
For upper \rgb\ stars with \logg $<$ \loggLimit, there is no need for a correction.

Finally, RC and HB stars have long been prized because of the narrow
range in absolute magnitude where they spend most of their lives
as CHeB stars. The Hipparcos catalog  \citep{perryman1997} confirmed the
theoretical prediction that stars have similar luminosities after
the helium flash in their CHeB phase (see, for example \citet{seidel1987}).
As a result, a pure sample of RC/HB stars can have distances inferred accurately --  an invaluable tool for mapping the Galaxy in stellar density and chemistry.
\cite{bovy2014} have provided such a list for several data releases of the APOGEE
\citep{majewski2017} survey. \cite{nidever2014} used the first of these
to show that distribution of stars in the [$\alpha$/Fe] vs. [Fe/H] space varies widely
across the disk of the Galaxy.
Recently, with the release of Gaia data \citep{gaia2016,gaia2018}, RC stars have been used to test for systematic errors in the reported parallaxes (e.g. \citet{davies2017}).

Clean separation between RGB and RC is vitally important but, as we indicated earlier,  can be difficult because the common observables
for stars: temperature, \logg (or luminosity), and [Fe/H] are degenerate for
red giants in the HR diagram. Stars with the same observables, but different interior structure
can overlap each other unless another parameter, for example, the mass of the star is also known.
\change{Where possible, we seek to identify which of the CHeB stars are  in the \rc .}
In all these areas of research, asteroseismology has a significant role.

This addition to the toolbox for stars with outer convection zones is provided by the natural resonances of the stars.
The Fourier spectrum formed from  time-series photometry shows clear evidence for the modes of oscillation of differing radial, latitudinal and azimuthal structure.
For evolved stars, all except the radial modes  are mixed modes which carry information from both the core
(gravity modes) and the stellar outer regions (acoustic modes).

We should note that data from the \textit{CoRoT} satellite were used to provide the  confirmation  of the existence of 
non-radial (and hence by definition mixed) modes in  the acoustic spectra of the red giants they observed.
For further details see \citet{deridder2009} and \citet{hekker2009} and references therein.
Soon after this, the extremely precise photometry from the {\it Kepler} mission \citep{borucki2010} of a large number
of red giants provided  new and revelatory data to probe the interiors of evolved stars.

From the early \textit{Kepler} data, \citet{bedding2010} reported the first clear detection of individual mixed modes in red giants.
The spacing of the mixed modes can be used to infer the evolutionary state of a star and a major step forward came when \citet{beck2011} measured these  spacings and then
\citet{bedding2011} used the typical spacing in period   between mixed modes to distinguish between RGB and CHeB stars for a large cohort of stars \citep[see also][]{mosser2011mm}.
The fundamental observational feature is that CHeB stars have a larger  spacing in period compared to that in RGB stars
\citep{dupret2009}.
These differences are partly caused by changes in the density differences between the core and the outer regions \citep{montalban2010}, as well as  the fact that in CHeB stars the core is (at least partly) convective \citep{jcd2014book}.
Following on from the identification of the evolutionary state of the star using the structure of the mixed modes came the discovery by \citet{kallinger2012} that the location of the radial modes is influenced by the evolutionary state of the star. This will be further discussed in Section~\ref{TK}.

As indicated earlier, using the seismic data together with  high quality spectroscopic data is a powerful approach to the determination of stellar properties.
For example, to obtain mass, radius and age information for a cohort of stars, one can use the combined observational data as input to stellar models.
Grid-search methods are used to determine (in a statistical
sense) the model that best fits the data (see \citet{rodrigues2017} and references therein).
The APOKASC project which uses \textit{Kepler} asteroseismic data and APOGEE spectroscopic data
is an  early application  of this combination of data and methods \citep{pinsonneault2014}.
However, a limitation with this previous work was that it did not include any information about the evolutionary state to the grid-based models.

For the next step in the APOKASC project, reported here,  we remedy this limitation.
This paper describes the processes that were followed to determine the evolutionary states of the individual stars. Other papers will describe their use in the determination of the characteristics of the individual stars.
In particular, the details of the spectroscopic analysis and an empirical correction to get reliable masses from the asteroseismic data, together with a catalogue of the relevant data,  is presented in \citet{pinsonneault2018}

We present several classification techniques, compare their results, and describe their
strengths and limitations. First, in Section~\ref{Observation}, we present the observational data  used.
Next, in Section~\ref{individualMethods} and Section~\ref{spectroscopic}, we describe the individual classification methods and then,
in Section~\ref{classification}, describe the methods used to  produce  the  consensus  values.
The results are given in  Section~\ref{detail} and an update to the classification results is given in Section~\ref{update}.
We finish with Discussion and Conclusions in Section~\ref{summary}.

\section{Observational Data}
\label{Observation}
We obtained  frequency power spectra from the timeseries observed during the {\it Kepler} mission
\citep{borucki2010} and near-infrared spectra from the APOGEE Survey
\citep{eisenstein2011,majewski2017} for thousands of red giants.
We refer to this combination as APOKASC.

\subsection{\textit{Kepler} Data} \label{kepler-data}
The photometric data from {\it Kepler} are sampled at an interval of approximately 30 minute (the so called
`long cadence' data). From these data, an acoustic spectrum is formed which can be characterized
by the frequency at which the oscillations are strongest, \nuMax, and the typical spacing,
\DeltaNu\ between pressure modes of the same degree, $\ell$, at successive orders.
These two quantities  are global seismic parameters.
The radial modes with $\ell=0$ are pure pressure modes, approximately equally spaced in frequency,
and nearly all the other, higher degree, modes have a mixed character being influenced both by buoyancy (g) and pressure (p).
The exception to this is where the influence of the buoyancy is  insignificant. 
In the so-called (first order) asymptotic expansion, the radial modes are equally spaced in frequency.
The underlying pure g modes are equally spaced in period.
The observed mixed mode pattern is therefore a mixture of features with quasi~regular period and frequency spacings.
Hence, one observes the underlying equal spacing in period slightly perturbed  where the g modes oscillate near an acoustic resonance.
More detail on this can be found in the review article by \citet{hekkerJCD2017} and the references therein.

\subsection{APOGEE Data}
For this project, stars in the \textit{Kepler} field were supplemented with
high quality spectroscopic parameters by APOGEE.
Part of SDSS-IV\citep{blanton2017}, the APOGEE survey for the
\textit{Kepler} field uses the Sloan Foundation Telescope \citep{gunn2006} at Apache
Point Observatory.
Further details on the target selection in APOGEE, including the selection of oscillating red giants in the
\textit{Kepler field}, can be found in \citet{zasowski2013}, \citet{pinsonneault2014} and \citet{zasowski2017}.

A multi-object H-band spectrograph  \citep{majewski2017,wilson2019}  takes moderate resolution (R$\simeq$ 22,500) spectra of up to 250 science targets at once.

After the data are reduced to 1-D spectra, with the usual steps such as flat-fielding and
wavelength calibration having been applied \citep{nidever2015}, the stellar parameters and
individual abundances are derived by ASPCAP \citep{garciaPerez2016}.
ASPCAP uses a comparison grid of synthetic spectra to report a $\chi^2$ minimum set of
`raw' parameters.
These raw values are then compared with accurate values for the subset of stars
with measurements based on fundamental techniques (for further information on this see \citet{holtzman2018}).
In particular, and importantly, the asteroseismic \logg\ is used to calibrate the spectroscopic gravities.

The uncertainties in the final spectroscopic calibrated values for  T$_{\rm eff}$, \logg, [Fe/H]
vary with stellar parameters and with S/N.
Median uncertainties in the APOKASC-2 sample are 76\,K, 0.05\,dex and 0.03\,dex respectively (Pinsonneault et al. 2018).
For a detailed discussion of the full APOGEE sample and approach see \citet{holtzman2018}.

We will later, in Section~\ref{specSigma}, return to the issue of the estimate of the statistical uncertainty on
T$_{\rm eff}$ when we compare the spectroscopic and seismic classifications.

\section{Individual Seismic Methods}
\label{individualMethods}
In this section we briefly describe the individual methods for determining the evolutionary states of individual stars and give the key features of the methods.
The characteristics of the methods are quite different which is a strength of the use of the four together to obtain consensus determination of the evolutionary state of individual stars.
The methods are numbered.

All the methods aim to separate first ascent RGB stars from RC stars. However, with the exception of Method~3, the methods are unable to distinguish between RGB and \agb\ (AGB) stars. 
There are particular difficulties to seismically identify AGB stars and to distinguish them from high luminosity RGB stars.
In both cases the \DeltaNu\ values are low, in general below the values seen for clump stars, and hence the \DeltaNu\ value on its own is not enough to allow the distinction to be made.
\citet{grosjean2014} discusses the issues surrounding the seismic observations of AGB stars.
As indicated in \citet{stello2013}, models suggest that the period spacing for AGB stars becomes  very similar to that of RGB stars.
Additionally, the very small period spacing is not observable with the data duration of the \textit{Kepler} data.
\citet{mosser2014} present observations of stars that are in transition between the clump and the AGB for which the period spacings are intermediate between the clump and the AGB values.
As stars evolve on the AGB and become more luminous, their evanescent zones become thicker and the coupling between the pressure and gravity modes is attenuated.
Consequently, the dipole modes progressively lose their mixed character and become essentially acoustic.
Consideration ought also be given to the large mode density \citep{mosser2018}.

\subsection{Elsworth $\equiv$ Method 1}
Method 1 \citep{elsworth2017evolYE}  is an autonomous  way of determining the evolutionary state from an analysis of the morphology of the power spectrum of the light curve.
The structure of the dipole-mode ($\ell=1$) oscillations, which have a mixed character in red-giant stars, is used to obtain  some measures that are used in the categorisation.
The feature of the structure that allows the algorithm to work is essentially that for RGB stars the period spacing is small and the mixed modes are relatively close to the location of the nominal p mode but for the CHeB stars the spacing is much wider.
The separation between \rc\ and \scl\ is based on  the \nuMax\ value and the estimate of the period spacing of the dipole modes. 
For a star to be classified as \scl\ the \nuMax\ value must be above \uHz{50}
and the observed period spacing is required to be above \PL.

\subsection{Hekker $\equiv$ Method 2}
Method 2 \citep{hekker2017svm} is based on grid-based modelling using the global asteroseismic parameters defined  in  Section~\ref{kepler-data}, 
\nuMax\ 
and \DeltaNu,
combined with effective temperature and metallicity.
In other words it aims to find loci in this four dimensional space where RGB and CHeB stars are uniquely defined.
To do this the authors use a C-type support vector machine with a Gaussian radial basis function as a Kernel.
In practice they perform the analysis on reduced parameter space where they use the ratio \DeltaNu/\nuMax .
This method has the advantage that it is applicable to relatively short datasets. The success rate of the method is of order 80 to 90\%.

\subsection{Kallinger $\equiv$ Method 3}
\label{TK}
Method 3 \citep{kallinger2012} uses established methods  automatically to locate and measure the frequencies of  the radial modes, i.e. those with degree of $\ell$ = 0.
From the frequencies is then determined the phase shift \eps\ of the central radial mode, i.e. the  offset in the linear, asymptotic fit to the acoustic modes.  They find that \eps, at a given \DeltaNu, is significantly different for RGB stars  which burn only H in a shell  and those that have already ignited core-He burning.

\subsection{Mosser $\equiv$ Method 4}
%
Method 4 is based on the measurement of the asymptotic period spacings $\DPi$, as given by \citet{vrard2016},
which relies on the asymptotic fit of the mixed-mode pattern \citep{mosser2012core}.
It uses  the methodology introduced by \citet{mosser2015} to stretch the oscillation spectrum in order to transform the varying period spacings of the mixed-mode pattern into uniform spacings equal to $\DPi$.
The Method-4 results used here were based on an early version of the method described.

As shown by \cite{vrard2016}, the method is not influenced by the presence of the signal due to the rotation of the star.
The method works on high-quality spectra characterized by a high signal-to-noise ratio, and requires that the value of the envelope autocorrelation function is greater than 100 in their units \citep{mosser2009}.
In order to locate precisely the  frequency ranges between radial modes where dipole mixed modes are observable, it requires the radial modes to be very  precisely identified.
This is done with the universal red giant oscillation pattern \citep{mosser2011up}.

The evolutionary stage classification follows the scheme depicted in \cite{mosser2014}: in short, RGB and RC stars are distinguished by the $\DPi$ value; stars in the \scl\ have a seismic mass above 1.85\,$M_\odot$.

\section{Spectroscopic Method $\equiv$ Method 5}
\label{spectroscopic}
The number of stars with sufficient lightcurve data for seismic analysis is considerably more limited than the number of stars with spectroscopy. 
For example,  $<$ 10\% of the stars with APOGEE spectra have {\it Kepler} or K2 (the re-purposed \textit{Kepler} mission) lightcurve data. 
Spectroscopic classification of evolutionary states based on temperature, surface gravity, and metallicity has a long history.
RC stars are hotter than RGB stars of the same gravity and metallicity, and they are  confined to a narrow range in surface gravity.
The secondary RC stars \citep{girardi1999}, which become a significant feature in intermediate-aged stellar populations, are also hotter than their RGB counterparts and have gravities that extend from the gravity of the RC to higher gravities. 
A single temperature does not cleanly divide the RC from RGB because of the sensitivity of the RGB temperature to metallicity and the shift of the mean RGB locus to higher temperatures at 
lower surface gravity 
\citep[see, for example, Figure~3 in][]{pinsonneault2018}.
Both the metallicity and surface gravity trends can be removed by comparing the observed temperature with a suitable reference temperature expected for an average RGB star of that gravity and metallicity.

\citet{holtzman2018} inferred the reference temperature empirically by fitting to the mean 
\Teff\ of the RGB stars in Pinsonneault et al. (2018)  as a function of metallicity and gravity, leading to a  reference temperature
\begin{equation}\label{eq1}
\rm T_{ref} = 4393.63 - 436.17 [Fe/H] +554.31 (log\textrm{(g)} - 2.5).
\end{equation}
The parameters in this expression are the uncalibrated versions.\\
The evolutionary state is then defined  as follows.

If \logg $<$ \loggLimit\ the star was classified as UpperRGB.
For such low gravities, the spectroscopic gravity corrections between the seismic and spectroscopic values are independent of evolutionary state.
This is not true for higher surface gravities and an important reason for the classification in APOGEE was to infer the correction to surface gravities.

If \logg $> 3.5$ the star was classified as DWARF, and asteroseismic detections are not expected in the long-cadence \textit{Kepler} data.
Across the high-gravity domain, the surface gravity correction was smoothly ramped to zero at \logg = \loggLimit.

For all other targets, the carbon-to-nitrogen ratio and the effective temperature are used to provide a dividing line between RGB and RC stars. The stars with the best-determined seismic states (i.e., where no pipeline returned a value of unclassified) from this work were plotted in the C/N vs. \Teff\ plane and a line placed to maximize the correct sorting of stars into RGB or RC.
A star is defined as RGB  if Equation~\ref{eq2} is satisfied and as RC otherwise.

\begin{equation}\label{eq2}
-256.41 \rm{[C/N]} > (\rm T_{eff} - T_{ref})  
\end{equation}

The physical justification for this relationship is as follows.
Although a simple temperature cut is effective, it does not account for mass trends in the RC and RGB locus.
More massive RGB stars are systematically hotter than lower mass RGB stars, making it more likely for them to be misclassified as RC stars.
Their lifetimes are also shorter, making them an uncommon population at higher initial mass.

RC stars have a double-valued locus (traditionally referred to in the horizontal branch as a candycane  diagram).
There is a faint lower branch, which becomes cooler with increased mass and reaches a minimum distance from the RGB.
It then transitions to a more luminous upper branch where higher mass stars are systematically hotter.

In the absence of detailed abundance information, these trends are difficult to account for.
However, the APOGEE survey can measure C and N abundances for large stellar samples, which is physically linked to stellar mass and metallicity.
The surface [C/N] ratio is modified during the transition from the main sequence to the \rgb\ by the first dredge-up \citep{iben1965II} which is mass dependent (see e.g. \citet{salaris2015Dredge}).
This can be used as a mass proxy \citep{martig2016, ness2016}, and it was employed as an additional tool for evolutionary state classification in SDSS Data Release 13 \citep{holtzman2018}.
The C/N mapping on to mass has intrinsic scatter, some of which comes from the effects of stellar mergers (see, for example, \citet{izzard2018}) and flattens at higher initial mass \citep{tautvaisiene2015} in open clusters.
We can therefore expect that using [C/N] as a mass proxy will improve spectroscopic classification, but also that there can be misclassifications.

An additional physical effect is the well-known phenomenon of extra mixing in metal-poor stars \citep{kraft1994}.
This was first seen in globular clusters as a secular decrease in C with increased luminosity on the \rgb\ of globular clusters,
which sets in above the \rgb\ bump (RGBB).
Because the luminosity of the RGBB overlaps  the luminosity of the RC, such mixing does not directly impact on the usage of C/N as a mass diagnostic on the lower RGB (but it does complicate efforts to infer the birth C and N as a function of metallicity).
Extra mixing will lower the surface C/N of metal-poor core-He-burning stars, which will be a bias in separating RC from RGB stars in that domain.
However this effect is mitigated in practice because the  metal-poor core-He-burning stars will tend to 
be outside the temperature domain probed by solar-like oscillations.
In the full APOGEE sample, extra-mixing is observed to set in for the oldest stars for [Fe/H]$ < -0.5 $ \citep{shetrone2019}.
Again, it is therefore plausible that this physical effect can lead to misclassification.

We will see that the comparison between spectroscopic and seismic classifications provides a way to get an estimate of the internal (statistical) uncertainties in the spectroscopic variables.
\section{How to form a consensus}
\label{classification}
In this Section we discuss the decision process used to produce the consensus (seismic) classification and then show the number of stars that fall into each class for both the seismic and the spectroscopic classification.
Through the remainder of the paper, the set of evolutionary state results produced by combining the different seismic classifications will be called the `consensus' set.

\subsection{Classifications provided by the different seismic methods}
In an ideal world, all the seismic methods would form the same \change{decision about }  the evolutionary state of a given star.
However, each of the methods have some level of mis-classification either due to the limitations of the method or to less-than-perfect quality data.
Furthermore, not all the methods provide the same detail in the classifications.
They all identify the  \rgb\ (RGB) and have an unclassified (U) category, but, at the time at which the classification was done for the APOKASC project, not all the methods distinguished between the different phases after the initiation of core-Helium burning, that is
\rc\ (RC) and \scl\ (2CL). In these cases, a classification of \textit{RC} should be interpreted as \rc\ or \scl.
There is also a level of difficulty in distinguishing between the \rgb\ and the Asymptotic Giant Branch (AGB) stars of similar luminosity.
The possible categorisations provided by each method are shown in
Table~\ref{individual:evol}.

We use the comparison between the different methods to improve the robustness of the evolutionary state assignments.

\begin{table}
\centering
\begin{tabular}{|c|c|}
\hline
Method&Designations \\
\hline
1&RGB, RC$\dagger$, U\\
2&RGB, RC$\dagger$, U\\
3&RGB, RC, 2CL, AGB, U\\
4&RGB, RC, 2CL,  U\\
\hline
\end{tabular}
\caption{The Evolutionary state designations provided by the different methods. Note that $\dagger$ indicates that the method does not distinguish between \rc\ and \scl\ stars.
For a fuller discussion of the terms see Section~\ref{classification}}
\label{individual:evol}
\end{table}

\subsection{Merging Values}
\label{merging}
The  evolutionary state determinations of the four different methods will be used together to produce a consensus value of the evolutionary state for every individual star.
In some cases the consensus value is entirely obvious because all the methods agree.
However, lower levels of agreement can still be useful but it is important for later application of these results that there is a record of the level of agreement in the  input values used.
This record is provided with the table of classifications available as an on-line document and detailed in the Appendix to this paper.

\begin{figure}
\includegraphics[width=0.5 \textwidth]{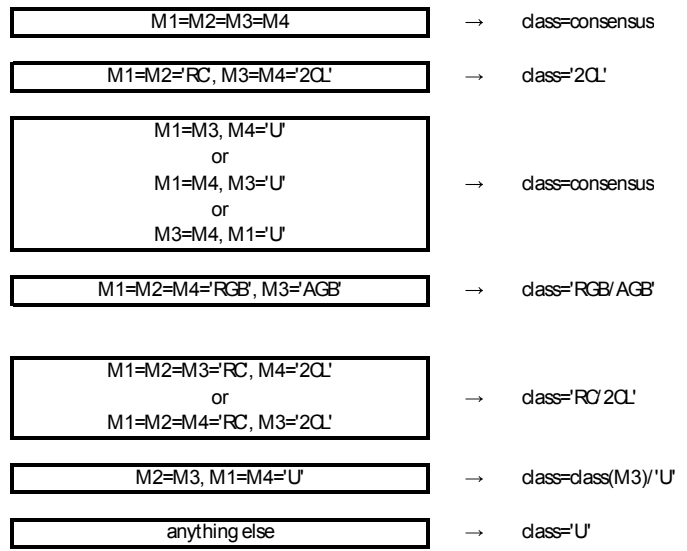}
\caption{The flow chart for the decision making process to form a consensus evolutionary state.}
\label{logic}
\end{figure}

We have  used ten different classifications, to be discussed shortly,  including ones which reflect some residual uncertainty in the evolutionary state of the star.
We next describe the process for forming a consensus and show a summary of the logic in Figure~\ref{logic}.

The first step in the process is  to check that there is agreement in the location in frequency space of the peak of the oscillation power (known as \nuMax).
Only if there is close agreement with the consensus value is the result of an individual  method used.
Some of the \textit{Kepler} spectra show evidence of more than one star and the presence of several stars will corrupt the determinations.
Simple mistakes or bad determinations of \nuMax\ are not normally relevant at this stage in the process and will  usually have been picked up earlier in the comparison process.

\medskip
In the determination of the consensus evolutionary state the following set of steps are followed.
\begin{enumerate}
\item  If all the designations  agree then the consensus classification is clear.

In the case of a \scl\ classification the situation is slightly different. Only methods~3 \& 4 produce this classification.
The equivalent in this case to `all agree' is that we require that methods~3 \& 4  give \textit{2CL} and methods~1 \& 2 give \textit{RC}.

There are other circumstances where we give a clear classification even though there is some disagreement between the classifications of the individual methods.
The first of these is where one of the methods fails to provide a classification, that is gives a classification of `U', or produces a classification which is different from that of the consensus. but  the remaining three methods agree with each other then the classification follows the views of the three that agree.

It has been recognised that method~2 has a higher mis-classification rate than the other three seismic methods. We therefore have a further level of determination of classification that ignores the classification provided by method~2. In this case, the requirement is that two of the remaining three methods agree and the third method does not return a value at all.

\item It is also possible to produce a classification that recognises an element of uncertainty between two evolutionary stages.
For AGB stars, the only method that produces this classification is method~3.
In this case,  if method~3 says \textit{AGB} and the other three methods say \textit{RGB} then we assign this star a classification of \textit{RGB/AGB}.
More work is needed with the seismic methods to learn how to distinguish between some of the RGB and AGB evolutionary states.

Similar logic is applied to the \textit{2CL} classification.
If only one of the two methods that can return a classification of \textit{2CL} does so but  all the other methods return a classification of \textit{RC} then we recognise the uncertainty and this star is classified as \textit{RC/2CL}.
There is no easy theoretical boundary between these two classifications and so it is not surprising that the seismic methods also find it difficult to draw the boundary.

\item
There is a final set of options that produce a classification but in this case the classification is partly uncertain.
The classifications concerned here are \textit{RGB/U, RC/U, 2CL/U, AGB/U}.
In this case neither method~1 nor method~4 provide any classification but method~2 and method~3 are consistent.
The classification then follows the value given by method~3.
Other combinations of methods are in principle possible but were not found to occur in the data.
\end{enumerate}

If all the methods indicate that they have failed to classify the evolutionary state of a star then the classification is \textit{U}.
There are several other situations that give rise to the star not having its evolutionary state classified and among these is the situation where there is an even split between different individual classifications assigned by the different methods.

\subsection{Statistics of agreement and disagreement}
\label{statistics}
We consider in this paper the so-called \textit{Second Summer} subset of the full APOKASC set.
The total number of stars considered  is \all.
Not all of the stars have been assigned an evolutionary state by all the methods and for some stars with very short datasets or for some that are probably not red giants, no classification is available.

Furthermore, as we use consensus between methods to arrive at a consensus evolutionary state, we do not provide a classification  where there is a single determination.

Out of the \all\ stars, the vast majority, that is 6060, have an assignment from all the methods, 478 from three, 77 from two, 28 from just one and 18 from none.

Furthermore, we recognise that there are limits to the \nuMax\ range where the data can be reliably processed.
Stars with \nuMax\ above \uHz{250}, but below \nuMax\ values typical of sub-giant stars, are undoubtedly RGB stars but we cannot trust their \nuMax\ values because of the proximity of the Nyquist frequency for the long cadence data.
We therefore do not classify them.
It is difficult to classify the stars with very low \nuMax\ because the time duration of the data, the small number of modes in the spectrum and the expected very low period spacing for these high luminosity  stars.
At the observed luminosity, they must be either RGB or AGB stars.
We have not put  a hard cut off here but many of these stars are unclassified because of the difficulties.

We next provide the summary statistics for the stars in the broad classifications of \textit{RGB}, \textit{RC} \& \textit{2CL}, and \textit{U}.
The stars that may be AGB are also considered.
The data are summarized in Table~\ref{agreed}.

\subsubsection{RGB stars}
First we consider  the RGB stars.
The number of stars that were classified as \textit{RGB} within the working range is 3372.
Of these 1996 (about 59\%) were given the same classification in all the methods.
A further 690 (about 20\%) were given the \textit{RGB} classification by three methods with the fourth method not returning a value.
As has been already noted,  method~2 does have a tendency to misclassify some of the stars and of the 490 (about 15\%) stars where one of the methods gives a conflicting classification, some 459 come from  method~2.
We now move to the situation where two of the methods plus an unclassified are used to determine the consensus evolutionary state and  method~2 is discounted.
Here a further 196 (about 6\%) of the stars are classified.

\begin{figure*}
\includegraphics[width=0.99\textwidth]{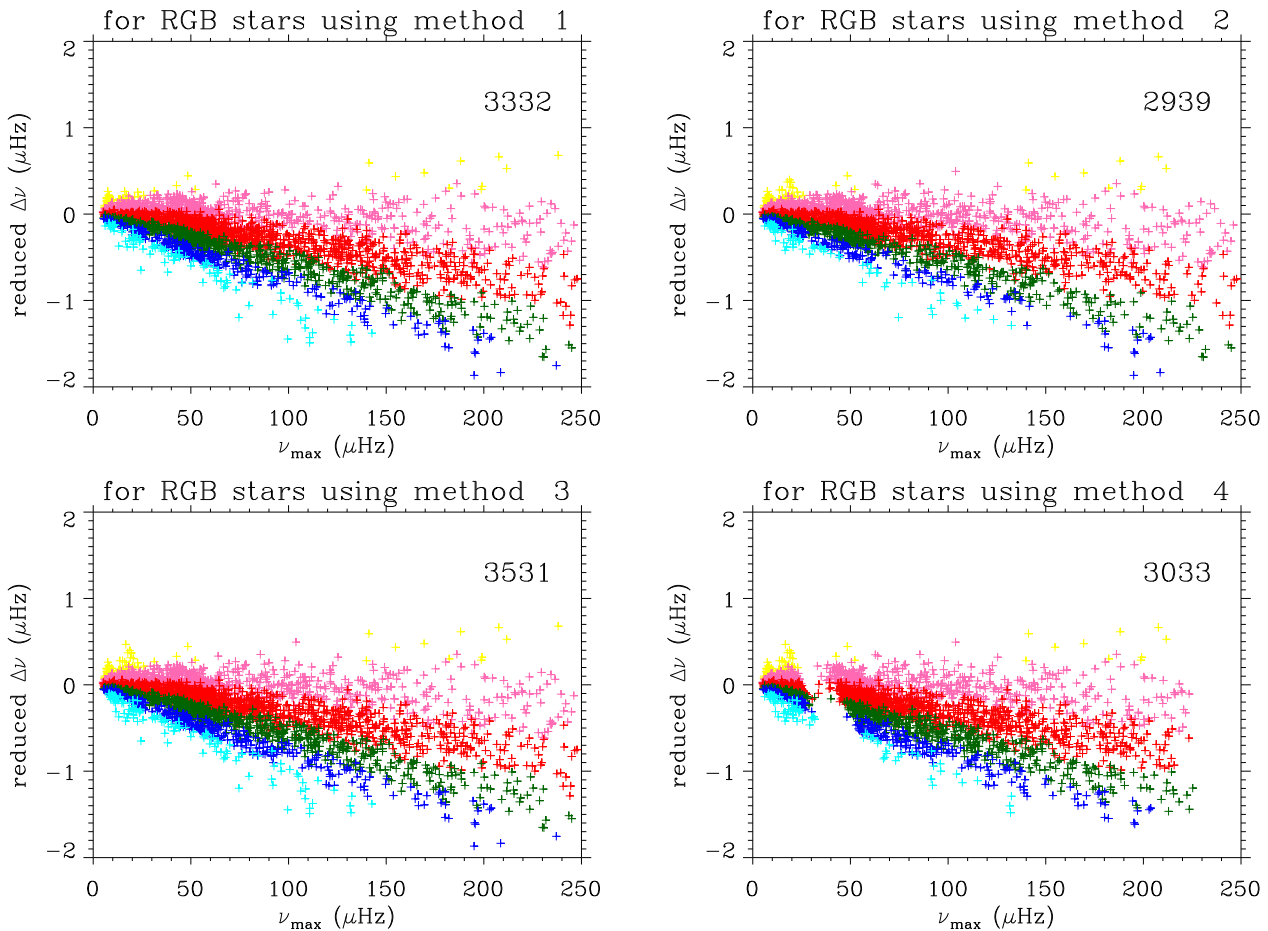}
\caption{The location in the reduced \DeltaNu\ vs. \nuMax\ plane of the stars whose evolutionary state for the given method is consistent with the `consensus' state which in this case is \rgb. 
Each panel is for a different method as indicated in the title. 
The number in the top right corner is the total number of stars in the panel. 
See the main text for a discussion on what is meant by `reduced \DeltaNu '.
The colour is indicative of the mass of the stars ranging from less than 1.0\Ms (in yellow) to more than 1.8\Ms (in cyan) in steps of 0.2\Ms.
Note that the gap in data for method~4 at around \nuMax\ = \uHz{40} 
is an indication of the difficulty that the method has in identifying RGB stars in this region.}
\label{RGB-agree}
\end{figure*}

In  Figure~\ref{RGB-agree} for RGB stars only we show, for each of the individual methods, the stars where the classification was provided and the consensus classification are consistent.
The data are shown on a \DeltaNu\ vs. \nuMax\ plot where the \DeltaNu\ value has been adjusted by subtracting a trend line for \DeltaNu\ given \nuMax .
The trend line chosen is:
\begin{equation}
\Delta\nu=a\nu_{\rm max}^b
\label{eq3}
\end{equation}
where a=0.254 and b=0.78. The precise values of these constants are unimportant; the purpose of the subtraction is to show the range of values in the data more clearly.
The difference between the trend line and the observed values is expected to be mass dependent with stars with higher masses expected to be at more negative values.
This trend is clearly visible in the plots which show the range  of \nuMax\ values considered 
from the low frequency end where the stars are clearly upper RGB to the high frequency end where the limitation is the proximity to the Nyquist frequency of the long-cadence \textit{Kepler} data.
We calculate the mass of individual stars using the scaling relations \citep{kb1995} involving \DeltaNu, \nuMax, and effective temperature. 
We recognise that corrections need to be applied before the mass values 
are accurate, but for illustrative purposes, the scaling-law mass formula is sufficient.
For a fuller discussion of the different approaches to correcting the scaling-law masses see \citet{hekker2019}.
We illustrate the mass effect by using colour in the plots to indicate the mass calculated as the seismic mass without any corrections.

\subsubsection{CHeB}
Next we consider the CHeB stars.
There are 1984 stars classified as \rc. Of these, 1563 (about 79\%)  were agreed upon by all four methods and 238 (about 12\%) were based on three methods agreeing plus one not returning a value.
Next, 152 (about 8\%) were based on three methods agreeing with each other and with one method actively disagreeing.
The remaining 31 (about 2\%) were based on disregarding the method~2 classification and looking for two other to agree and the third not returning a value.

\begin{figure*}
\includegraphics[width=0.99\textwidth]{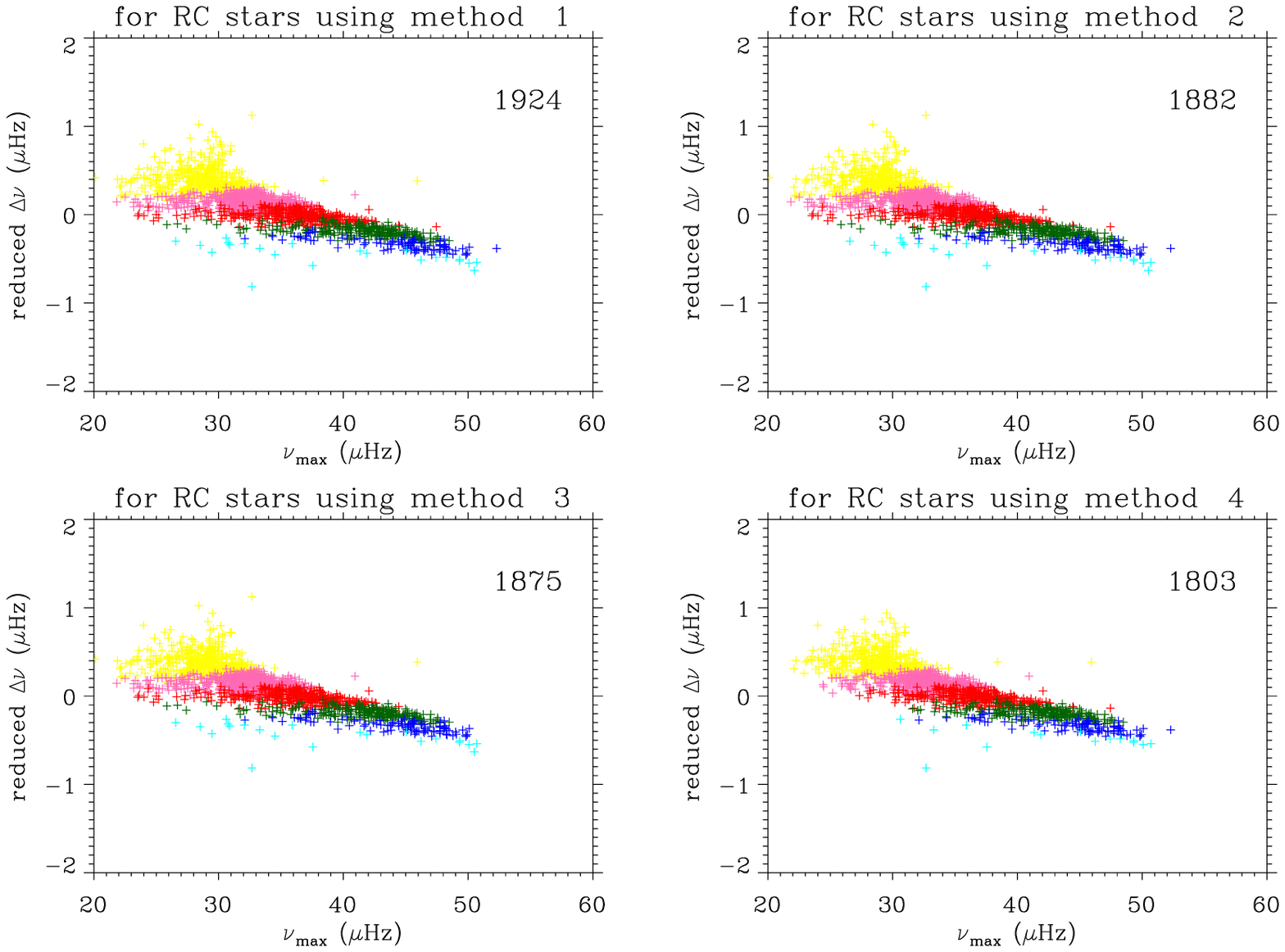}
\caption{The location in the reduced \DeltaNu\ vs. \nuMax\ plane of the stars whose evolutionary state for the given method is consistent with the `consensus' state which in this case is \rc. Each panel is for a different method as indicated in the title. The number in the top right corner is the total number of stars in the panel. See the main text for a discussion on what is meant by `reduced \DeltaNu '.
The colour is indicative of the mass of the stars ranging from less than 1.0\Ms (in yellow) to more than 1.8\Ms (in cyan) in steps of 0.2\Ms.}
\label{RC-agree}
\end{figure*}

In  Figure~\ref{RC-agree} for RC stars  we show, for each of the individual methods, the stars where the classification was provided and the consensus classification are consistent.

The range of \nuMax\ shown reflects the values at which CHeB stars exist.
It can be seen from these figures that the performance of the different methods is very similar. There are some differences in the range of coverage but they are small.

We have attempted to distinguish between \textit{\rc} and \textit{\scl} in the consensus classification, however, not all the methods made the distinction.
We  report here  the total number of Secondary-Clump stars (=286) which is made up of all the cases where there is  agreement (=233) among all the relevant methods,  and the cases where only one method disagrees (=53). This latter number is not reported in Table~\ref{agreed}.

For 179 stars the classification was uncertain between \textit{RC} and \textit{2CL}.
For the purpose of the grid-based modelling for which these data were prepared, the RC/2CL distinction was not considered important.

\subsubsection{Other classification}
Only one method was able to classify AGB. In order to have a consensus classification of \textit{RGB/AGB} that allowed for the possibility that the star was an AGB, it was required that method~3 returned \textit{AGB} and the others returned \textit{RGB}. There were 260 such instances.

There are 116 stars with a partially uncertain classification the vast majority of which are probably core-Helium burning stars based on a visual inspection of their acoustic spectra.
The remaining stars were totally unclassified. We will come back to these in Section~\ref{detail}.

In total, 6197 (93\%) are given some level of evolutionary state classification and 464 (7\%) are unclassified.
As is obvious from the numbers given here and in Table~\ref{agreed}, there is a high level of agreement between the different methods.

We next look for any trends in the conditions in which there is agreement and/or disagreement between the methods.
\begin{table}
\begin{tabular}{|l|c|c|c|c|c|}
\hline
Agreed&All&4 agree&3U&3X&2* \\
\hline
RGB&3372&1996&690&490&196\\
RC&1984&1563&238&152&31\\
2CL&286&233&-&-&-\\
RC/2CL&179&-&-&-&-\\
RGB/AGB&260&-&-&-&-\\
\hline
RGB/U&7&-&-&-&-\\
RC/U&88&-&-&-&-\\
2CL/U&20&-&-&-&-\\
AGB/U&1&-&-&-&-\\
\hline
U&464&-&-&-&-\\
\hline
\end{tabular}
\caption{Number of stars in each category where there is a consensus classification. The first column is the given classification as discussed in Section~\ref{merging}.
The second column (All) is the total number of stars given the categorisation. The third column (4 agree) is the numbers of stars where all methods agree.
The fourth column (3U) is where 3 methods agree and the other method does not provide a classification.
The fifth column (3X) is where 3 methods agree and the other method disagrees.
The final column (2*) concerns the case where two out of methods~1, 3 \& 4 agree and the other of these three methods does not provide a classification.
For completeness, the final line in the table gives the number of unclassified stars. }
\label{agreed}
\end{table}

\subsection{Visualization of stars with firm classifications}
\begin{figure*}
\includegraphics[width=0.99\textwidth]{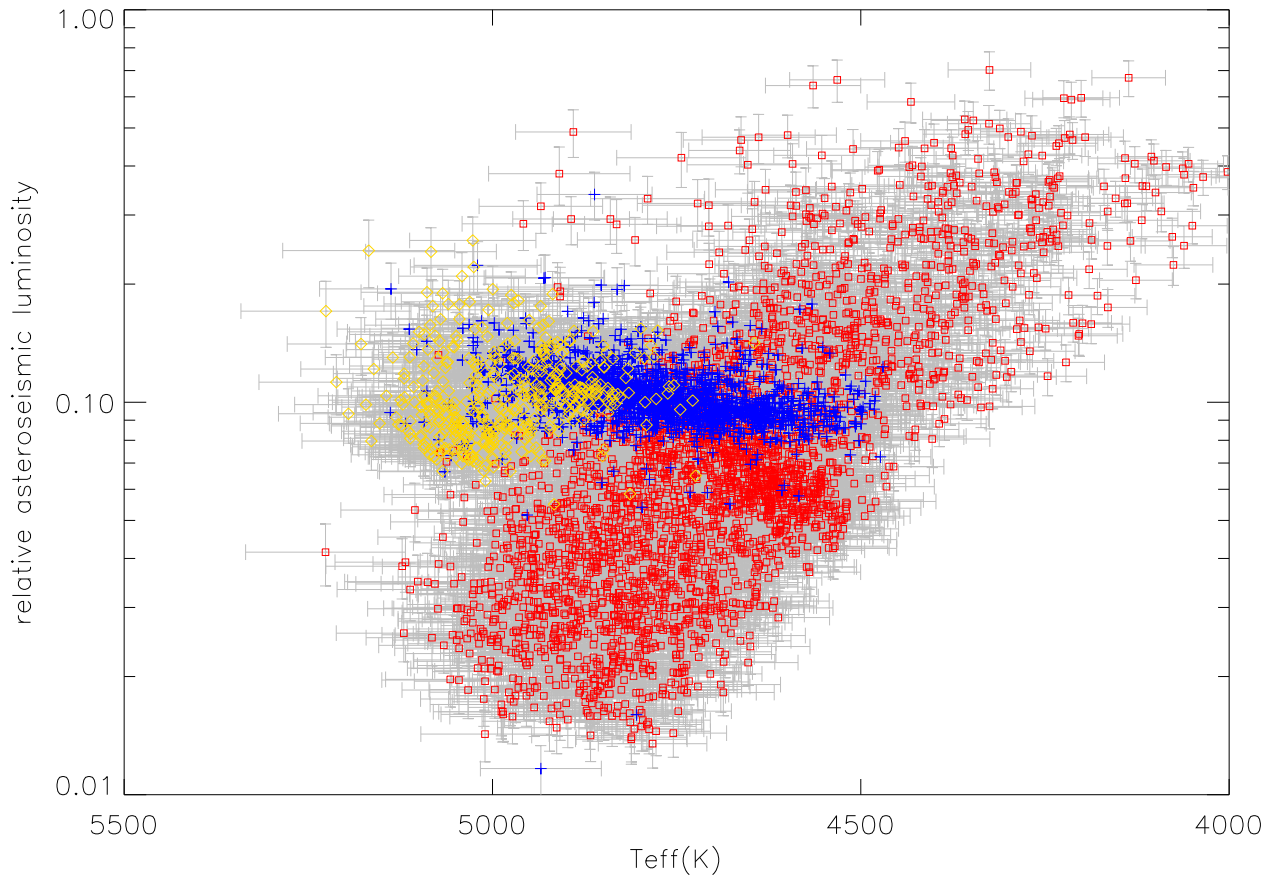}
\caption{An asteroseismic Hertzsprung-Russell diagram for the stars with an asteroseismic classification. Indicative uncertainties are shown in grey. RGB and RGB/AGB stars are shown as red open squares, RC stars as blue crosses and 2CL and RC/2CL stars are yellow open diamonds.}
\label{HRD-agree}
\end{figure*}

As a validation of the results obtained, we consider the classical luminosity vs. temperature Hertzsprung-Russell diagram (HRD) for the stars in this study.
Figure~\ref{HRD-agree} shows an asteroseismic HRD for the stars with an asteroseismic classification. In this plot the  RGB and RGB/AGB stars are shown as red open squares, RC stars as blue crosses and 2CL and RC/2CL stars are yellow open diamonds.
The asteroseismic luminosity, $L$, is calculated from the global seismic parameters and effective temperature using the relationship 
\begin{equation}\label{eq4}
L=\frac{\nu_{\rm max}^2}{\Delta\nu^4}T_{\rm eff}^5
\end{equation}
The numbers are then divided by $10^{20}$ to give the range shown in the plot.
The uncertainties shown in grey are only indicative of the likely uncertainty. Instead of the full uncertainty on the effective temperature we could have chosen the somewhat lower likely statistical uncertainty which we will determine when the comparison is done between the seismic and spectroscopic classifications in Section~\ref{detail}.

The RGB and CHeB stars are located where one would expect them to be and the relatively low uncertainties plus the small number of outliers are an additional mark of the quality of the data. 

\subsection{Reasons for difficulty in seismic classification}
\label{unclassified}
The \textit{Kepler} data used in this analysis are taken with a sample time of close to 30\,minutes which means that the Nyquist frequency of the spectrum is about \uHz{283}.
Stars that have the oscillation power at a frequency near to the Nyquist frequency (called high \nuMax\ stars) suffer from the problem that some of the oscillation power is reflected about  the Nyquist frequency leading to a distorted spectrum.
For stars with \nuMax\ above the Nyquist frequency, the spectrum  appears at the wrong frequency having been reflected down from higher frequencies.
We find that some of the stars at high \nuMax\  values are spectroscopically classified as
DWARF.
This is entirely reasonable and is one way of identifying these super-Nyquist stars.

At low frequency, the number of modes visible in the spectrum gets quite small and the mixed character of the dipole modes is often not seen.
It is possible here to use some of the methods to distinguish between RGB and AGB stars but the options become more limited.

Furthermore, there are a few general reasons why stars are quite likely to fail to be classified seismically. These reasons are (a)~a very short data set; (b)~very poor duty cycle (or fill) in the time series;  (c)~little power in the dipole modes; (d)~the proximity of another star to the one under study causes contamination of the light curve.
This is not to say that these stars cannot be classified but it does indicate that there may be difficulties. We next look at these options  in more detail.

 We  have to remember that although there are many very long data sets obtained with the
\textit{Kepler} satellite, there are others that are very short, and furthermore, some of the stars fell, periodically, on detectors which ceased to function during the  mission.
There are other causes of breaks in the datasets due to issues with the spacecraft itself and decisions about whether or not to observe a given star.
This means that there will be stars with a significant time interval between the first and the last data point and with periodically interrupted time coverage and hence less than 100\% duty cycle.
Stars with very short data sets may well have 100\% duty cycles.
In short, we need to consider both duty cycle and the number of good data points when deciding when the spectrum is likely to be suitable for classification purposes.

We find empirically, that a fill of less than 50\% and/or  a number of good data points equivalent to an overall data duration of less than 50 days tends to make it difficult to have an consensus classification for a star.
A similar conclusion was reached by \citet{mosser2018}.

In summary, although we emphasise that we do not consider the duty cycle of the data when determining the consensus evolutionary state of a given star, we are not surprised when a star with duty cycle less than 50\% is unclassified.
Similarly we consider that data durations of less than 50 days are unlikely to produce clear classifications with the current methods.
In the light of upcoming missions like TESS \citep{ricker2014}, new and/or improved classification methods will have to be developed. We discuss later in this paper some of the new methods which may be successful with shorter data sets.

The other source of difficulty for the individual classification methods is where the dipole modes are very weak. Two of the four methods rely on the structure of the dipole modes.
If the dipole modes are largely absent, their characteristics cannot be used.
We can therefore expect this to be a reason why the classification might fail.
We will come back to the possible origins of this phenomenon in  Section~\ref{summary}.

Contamination of the light curve supposed to be of a single star with light from another star can come from chance alignments of the stars or from two (or more) stars that are physically associated with each other.
The presence of light from more than one star in the \textit{Kepler} aperture can be detected in various ways.
The fractional strength of the granulation and oscillations are a function of \nuMax\ and so are predictable \citep{mathur2011}.
If there is more than one star in the aperture, then, when  the data are scaled to fractional intensity, the strength of the oscillations in the red-giant star will be attenuated.
In other words, when the data are contaminated, the  signal levels are lower than predicted and the noise levels are usually elevated.

Sometimes there is evidence  in the spectrum of two (or more) different spectral regions exhibiting oscillations.
The extra signal may be due to  another red giant, but it often comes from a classical pulsator.

There is a set of stars, extensively studied in \citet{colman2017}, for which the spectrum of the red giant is contaminated by  very strong spikes.
Some of these are due to chance alignments of more than one star in the field of view and some are believed to be physically associated.
Details of possible processes to explain the spikes are discussed in  that paper and are not considered further here.

It is also recognized that stars in true binary systems can have the strength of their oscillation attenuated \citep{gaulme2014}. These stars may be clearly detected as red giants from their granulation characteristics but it may be difficult to classify their evolutionary state.

\subsection{Spectroscopic classifications summary}
\label{unclassified-spect}
By the nature of the methods employed, the classifications produced by the spectroscopic method are somewhat different from those produced by the seismic classification.
The classifications used that are relevant to the red giants are \textit{RGB}, \textit{upperRGB}, \textit{RC}, 
There is also a \textit{DWARF} category.
For the precise meanings of these terms see Section~\ref{spectroscopic}.
\begin{itemize}
\item 3108 stars are classified spectroscopically as being \textit{RGB} and 693~\textit{upperRGB}.
\item In the post RGB phase there are 2840~\textit{RC}.
\item There are 21 \textit{DWARF} stars.
\end{itemize}

The APOGEE pipeline solves simultaneously for the parameters used in the spectroscopic evolutionary state classification (T$_{\rm eff}$, [Fe/H], \logg, [C/H], and [N/H].)
As long as the pipeline can find solutions with an acceptable $\chi^2$ value for all parameters it is possible to derive a spectroscopic evolutionary state.
For a small number of targets, however, the final fit in one or more of the parameters listed above is poor, and APOGEE does not provide stellar parameters in these cases.
Some 3 stars in the APOKASC-2 sample had poor C or N data, while 13 more had multiple bad stellar parameters.

\section{Seismic and Spectroscopic viewpoints compared}
\label{detail}

All the stars have been considered by both the spectroscopic and seismic analysis although for operational reasons some stars will be missing from one or other classification.
The reasons  have been presented in section~\ref{unclassified} for the seismic methods and in section~\ref{unclassified-spect} for the spectroscopic method.
Hence there is a small difference between the number of stars in the APOKASC-2 list and those with asteroseismic classifications.

In this section we will discuss the results of a comparison  between the spectroscopic and seismic evolutionary state classifications.
The results of this comparison are a key result of this paper.

We provide lists of the consensus evolutionary state together with the spectroscopic determinations  in the on-line version of the paper.

First we look at the spectroscopic uncertainties.
Although the consideration of them  is perhaps a small point, we discuss the issue first so that we can take it into account in the subsequent section where we go on to look at the level of agreement/disagreement between the different approaches.

\subsection{Spectroscopic Uncertainties}
\label{specSigma}

As indicated earlier, the spectroscopic division between RC and RGB in [C/N] vs. $\Delta$T space (where $\Delta \rm T \equiv \rm T_{eff} - T_{ref}$) was defined by the seismic sample. 
When the dividing line was drawn, it was clear that there were a few stars that would be spectroscopically defined incorrectly, in part because of statistical uncertainties in the spectroscopic parameters. 

We can use the disagreement between the spectroscopic and seismic classifications to get a handle on the statistical errors in the spectroscopic parameters.
The dominant source of statistical error in the spectroscopy is due to uncertainty in the temperature determination.
We find that the vast majority of the disagreement between the methods can be explained by a 44\,K uncertainty in the effective temperature of the RGB stars and a 39\,K uncertainty for the CHeB stars.
These values say nothing about any systematic uncertainties.
There are a few cases of stars for which the disagreement in the classification cannot be explained by temperature uncertainty.

\subsection{The level of agreement between seismic and spectroscopic classifications}

In general the level of agreement between the spectroscopic and seismic  approaches is significantly better than 90\% giving us confidence in the process.
While we recognize that in some cases the disagreement will be due to measurement uncertainties, we can also expect to find some stars that are  genuinely unusual and interesting.

\begin{figure*}
\includegraphics[width=0.99\textwidth]{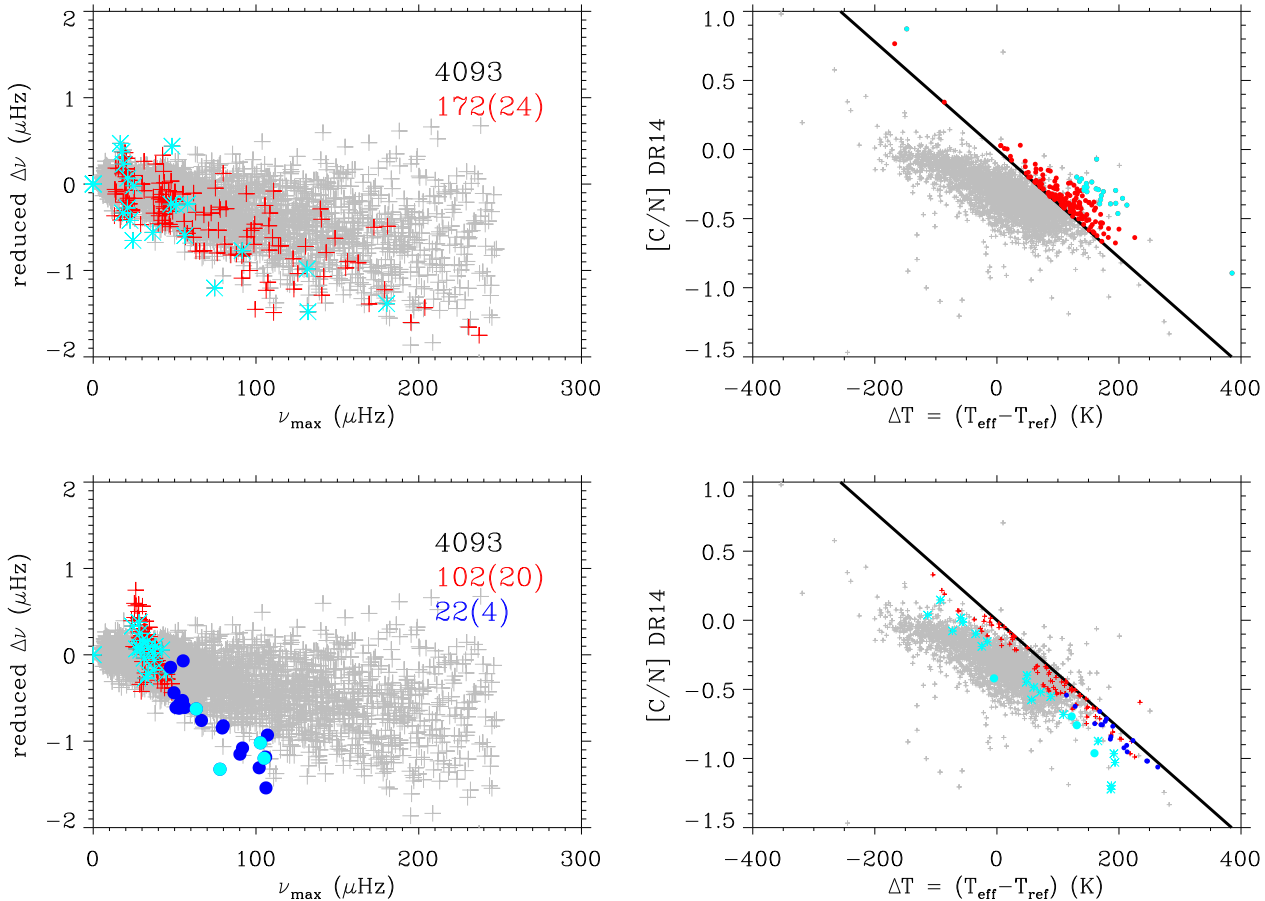}
\caption{For RGB evolutionary classification, a presentation of the agreement and disagreement between the consensus seismic and spectroscopic methods.
In the upper row are all the stars for which the consensus seismic evolutionary state is RGB.
The stars plotted in grey are those stars for which the spectroscopic method agrees.
In the lower row are  all the stars for which the spectroscopic evolutionary state is RGB.
The stars plotted in grey are those stars for which the consensus seismic method agrees.
In all cases, the red, blue and cyan features indicate where there is disagreement between the approaches.
For more detail and an explanation of the numbers inside the panels see the text \change{and Equation~\ref{eq1}.}}
\label{rgbContamination}
\end{figure*}

We are interested in the seismic RC and 2CL contamination within the spectroscopic RGB and vice versa.
This is shown in Figure~\ref{rgbContamination}.
In all cases the stars plotted in grey correspond to the stars that are classified as being in the RGB evolutionary state by both the seismical consensus method and the spectroscopic method
(method~5).

To gain insight into why stars might be mis-classified, we plot the data twice, once as we would view it seismically and once as we would view it spectroscopically.
In the left hand panels the axes give the location in the reduced \DeltaNu\ vs. \nuMax\ plane as previously discussed (see Equation~\ref{eq3}).
In the right hand panels, the axes give the location in the spectroscopic plane as  previously discussed (see Equation~\ref{eq2}).
There  is no spectroscopic \scl\ classification.


The discrepant  stars are shown as red crosses and cyan asterisks the upper, left-hand panel of Figure~\ref{rgbContamination}.
The red crosses indicate stars with spectroscopic uncertainties within $3\sigma$ of the RGB boundary.
There is hence some uncertainty in their RC designation and we may chose to exclude them from consideration as misclassified stars.
The cyan asterisks indicate stars that are outside  $3\sigma$ uncertainty.

The numbers shown in the upper left panel indicate the total number of stars for which there is agreement between seismic and spectroscopic methods (=4093) and underneath that the number of stars for which there is disagreement (=172).
In brackets is the number of stars for which there is disagreement and the spectroscopic value is outside the uncertainty (=24).



\begin{figure*}
\includegraphics[width=0.99\textwidth]{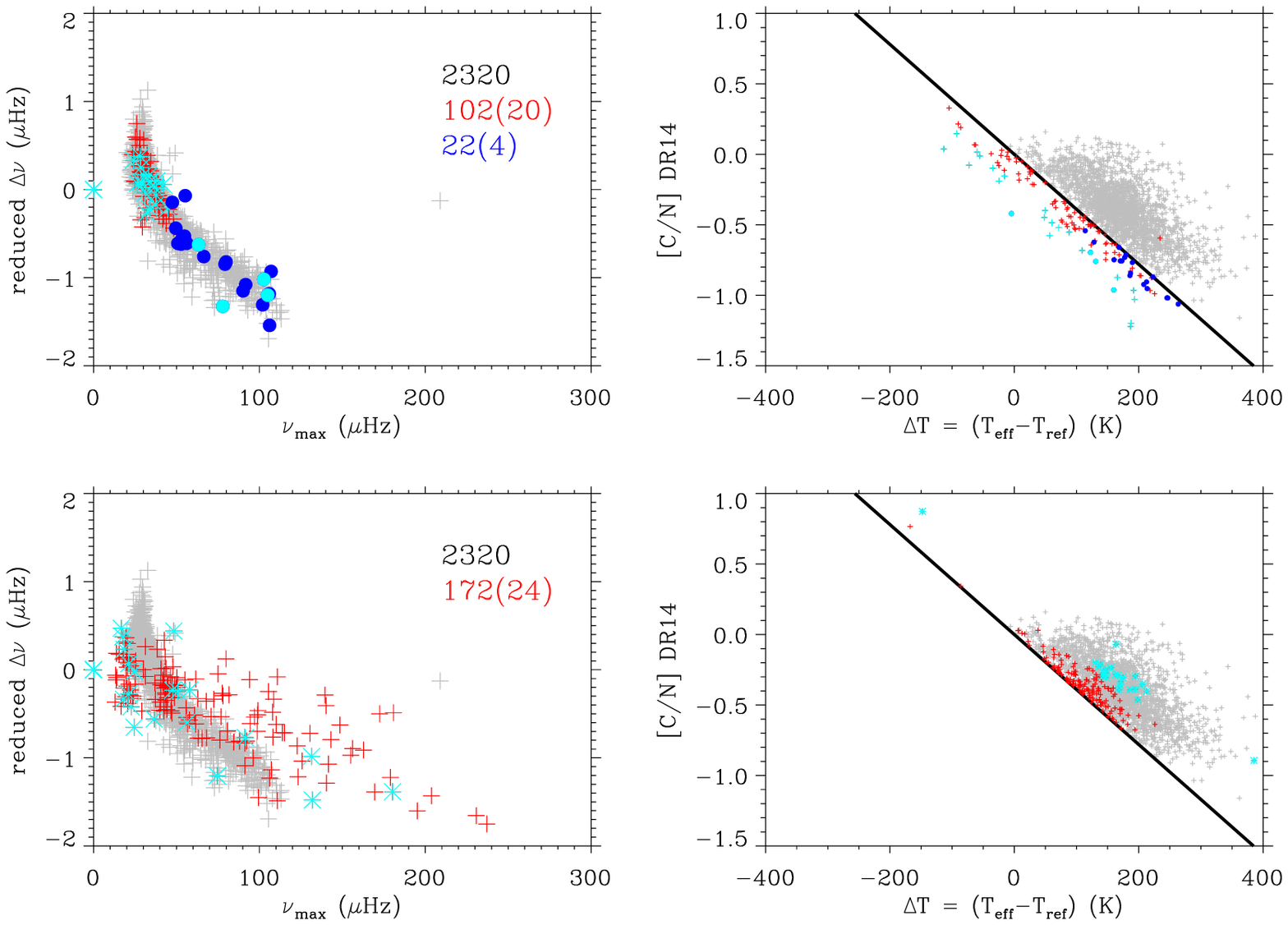}
\caption{For CHeB evolutionary classification,
a presentation of the agreement and disagreement between the consensus seismic and spectroscopic methods.
In the upper row are all the stars for which the consensus seismic evolutionary state is CHeB.
The stars plotted in grey are those stars for which the spectroscopic method agrees.
In the lower row are  all the stars for which the spectroscopic evolutionary state is CHeB.
The stars plotted in grey are those stars for which the consensus seismic method agrees.
In all cases, the red, blue and cyan features indicate where there is disagreement between the approaches.
For more detail and an explanation of the numbers inside the panels see the text  \change{and Equation~\ref{eq1}}.
}
\label{rcContamination}
\end{figure*}

We now consider the  stars which have a spectroscopic classification of RGB but the seismic classification is different.
These are shown in the lower row in Figure~\ref{rgbContamination}.
In this case we have to consider both \rc\ and \scl\ stars.
The \rc\ stars (shown as red crosses) tend to have lower masses than the \scl\ stars (shown as blue dots).
The diagonal track of these stars is quite unlike the distribution of the RGB stars and
(as will be obvious in Figure~\ref{rcContamination}) is much more like the track expected of \rc\ or \scl\ stars.

The notation for the numbers is the same as in the upper panel except that red is for seismic RC and blue is for seismic \scl.
The stars shown in cyan are those which lie outside the likely uncertainty range and are therefore mis-classified.

What conclusions can be draw from this analysis?
The first point is that there is a very high level of agreement between the methods.
Consider the RGB stars first.
As shown in the upper left panel of Figure~\ref{rgbContamination}, for RGB there are 2789 stars for which the classification agrees and only 172 stars for which seismic=RGB \& spectroscopy=RC (about 6\% of the total).
For the stars classified as RGB by spectroscopy, there are (by definition) the same number of stars where the seismic and spectroscopic approaches  agree and only
123 stars (about 4\% of the total) for which there is disagreement.

In summary, out of the \all\ stars, some 19 stars lack a spectroscopic classification, 464 stars lack a seismic classification and
only 1 star has a complete lack of classification in either system.
Of the stars without a seismic classification, 97 are ones that lie outside our classification system.
This is due to a variety of reasons; for example, bad time series, \nuMax\ close to, or above the Nyquist frequency, \nuMax\ very low and, in some cases, contamination by a classical oscillator in the field.
It is outside the scope of this paper to go further into this.
The lack of a spectroscopic classification is usually some operational difficulty as indicated in Section~\ref{unclassified-spect} and is not related to the characteristics of the individual star.

\subsection{Why the spectroscopic categorization might be wrong}
In considering the disagreement between spectroscopy and seismology the first thing to note is that the seismic methods are more fundamental because they are based on observations of the stellar interior.

Because the spectroscopic method was based on dividing the sample in the [C/N] vs. $\Delta$T plane to maximize the correct classification of the bulk populations, stars that are outliers in their population properties will be more frequently misclassified spectroscopically. 
These include stars with masses greater than $\approx$ 1.8\Ms, when the [C/N] value saturates at its lowest value regardless of mass (e.g. \citet{hasselquist2019}). 

The `young' alpha-rich stars that more massive than other thick disk stars are also problematic.
Many of these  stars  have  [C/N] ratios that are consistent with those of low-mass stars, indicating that they are in binary systems where mass was transferred after first dredge-up (e.g. \citet{jofre2016} and \citet{izzard2018}).
Consequently their surface [C/N] ratios no longer reflect the true mass of the star though they are still considered a reliable indicator of the true age of the star \citep{hekkerJohnson2019}

Finally, the division of the [C/N] vs. $\Delta$T with a linear relationship is probably too simple. 
However, using a more complex formulation  based on several slopes increases the risk of something going wrong  and has not been implemented for the APOGEE sample.
In the next subsection we will consider the stars that lack an asteroseismic classification.

\begin{figure*}
\includegraphics[width=0.90\textwidth]{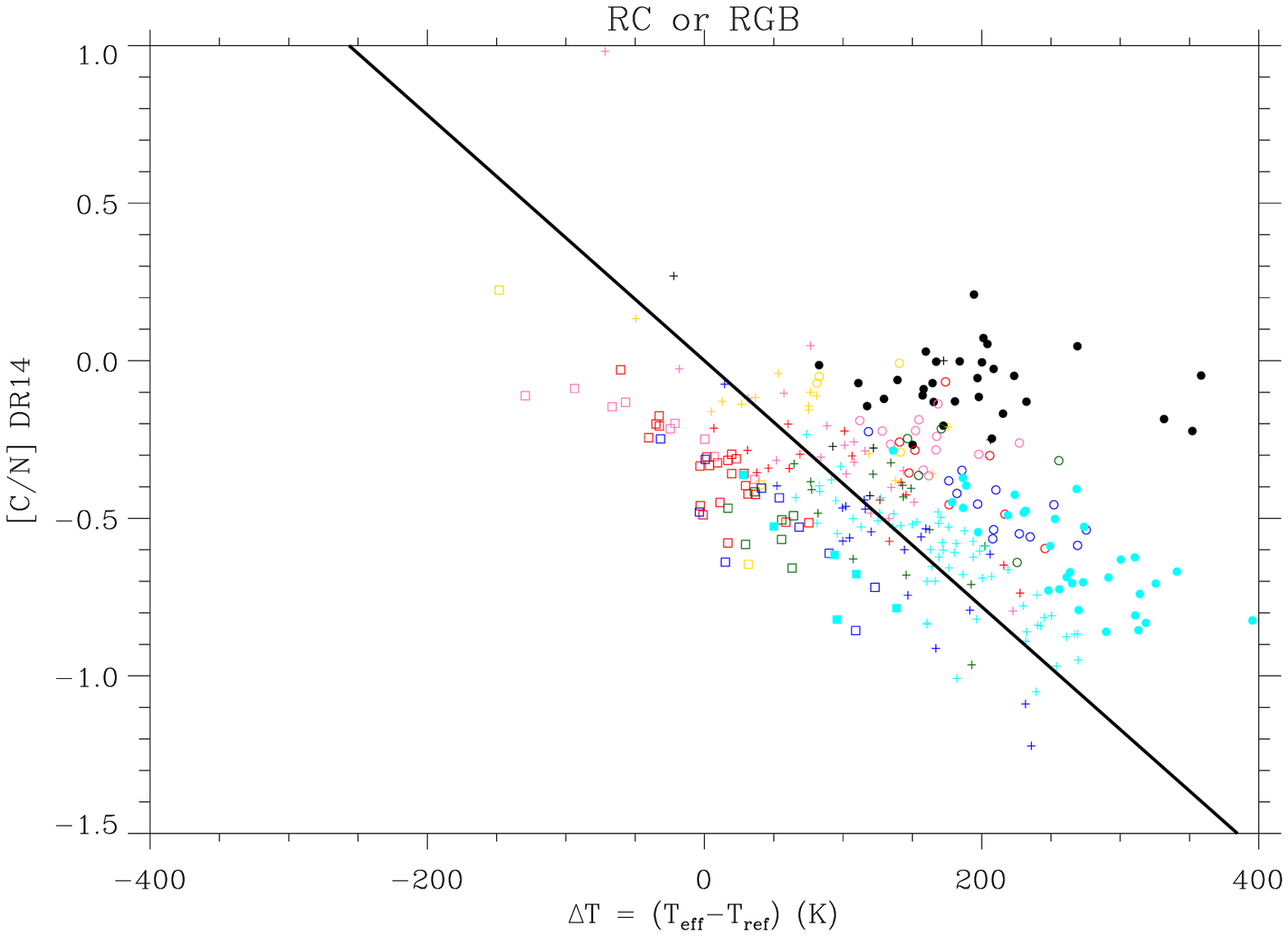}
\caption{The seismically unclassified stars for which there is a spectroscopic classification 
of either RGB or RC are plotted in the spectroscopic parameter space.
The spectroscopic UpperRGB classification is excluded.
The data are colour-coded by mass.
Black=0.5 to $<$0.8\Ms; yellow=0.8 to $<$1.0\Ms; pink=1.0 to $<$1.2\Ms;
red=1.2 to $<$1.4\Ms; green=1.4 to $<$1.6\Ms; blue=1.6 to $<$1.8\Ms; cyan=$>$1.8\Ms.
The small crosses indicate that the spectroscopy cannot make a clear choice between the classifications of RGB or RC. The squares (filled or open) indicate spectroscopic RGB and the circles (filled or open) indicate RC. For clarity, the extreme ends of the mass scale are shown as filled symbols.}
\label{seismicU}
\end{figure*}

\subsection{Seismically unclassified stars}

In Figure~\ref{seismicU} we show the spectroscopic parameters of the seismically unclassified stars where the colour is indicative of the seismic mass of the individual stars.
The clear mass trend observed is consistent with the spectroscopic classification method.

It can be seen that the stars are located in both the RC and the RGB domains.
It is also notable that there is a much more even spread of the masses across the  mass scale than is seen for the seismically classified stars.
Quantitatively, for seismically classified RGB stars about 2\% of the stars have masses below 1.0\Ms\ and about 3\% above 1.8\Ms;
for CHeB stars there are about 25\% below 1.0\Ms\ and about 1\% above 1.8\Ms.
However, for the seismically unclassified stars the numbers are about 19\% below 1.0\Ms\ and about 31\% above 1.8\Ms.

A detailed analysis of the reasons why each of the stars is seismically unclassified is beyond the scope of this work but there are a few features that we can comment on. 
For the stars with masses below 0.8\Ms\ the spectroscopic classification is overwhelmingly RC. 
These stars are discussed in Section~\ref{summary} and an illustration of their spectra is given in Figure~\ref{low-mass}. From the appearance of their seismic spectra, it is not surprising that seismic classification can be difficult.
The high-mass, seismically unclassified stars are also predominantly spectroscopically RC. 
This is probably indicative that the stars are in the \scl.

\section{Updates to the classification \change{methods}}
\label{update}
As indicated earlier in the paper, this project to produce evolutionary state classifications was part of the APOKASC programme to produce wide ranging information about red-giant stars.
The classifications reported here have been used as training sets for further methods, and have allowed the refinement of existing methods.

We will very briefly mention one new method here  from \citet{hon2018} who applied machine learning to the problem.


There are many approaches that can be taken for how to compare methods. Here we concentrate on a couple of key issues.
Firstly, we would like to see a reduction in the number of unclassified stars in order to provide as wide a sample as possible.
The second  key issue refers to the classification of stars as being in the \rc.
Reliable identification of \rc\ stars is important because, as discussed in the introduction, they can be used as distance indicators.
If that sample is contaminated with either \rgb\  or \scl\ stars then this dilutes the usefulness of the sample.
It is relatively easy to exclude the \scl\ stars by reason of their relatively high masses and their \nuMax\ values.
What is much more serious is contamination of the sample with RGB stars.

\begin{figure}
\includegraphics[width=0.49\textwidth]{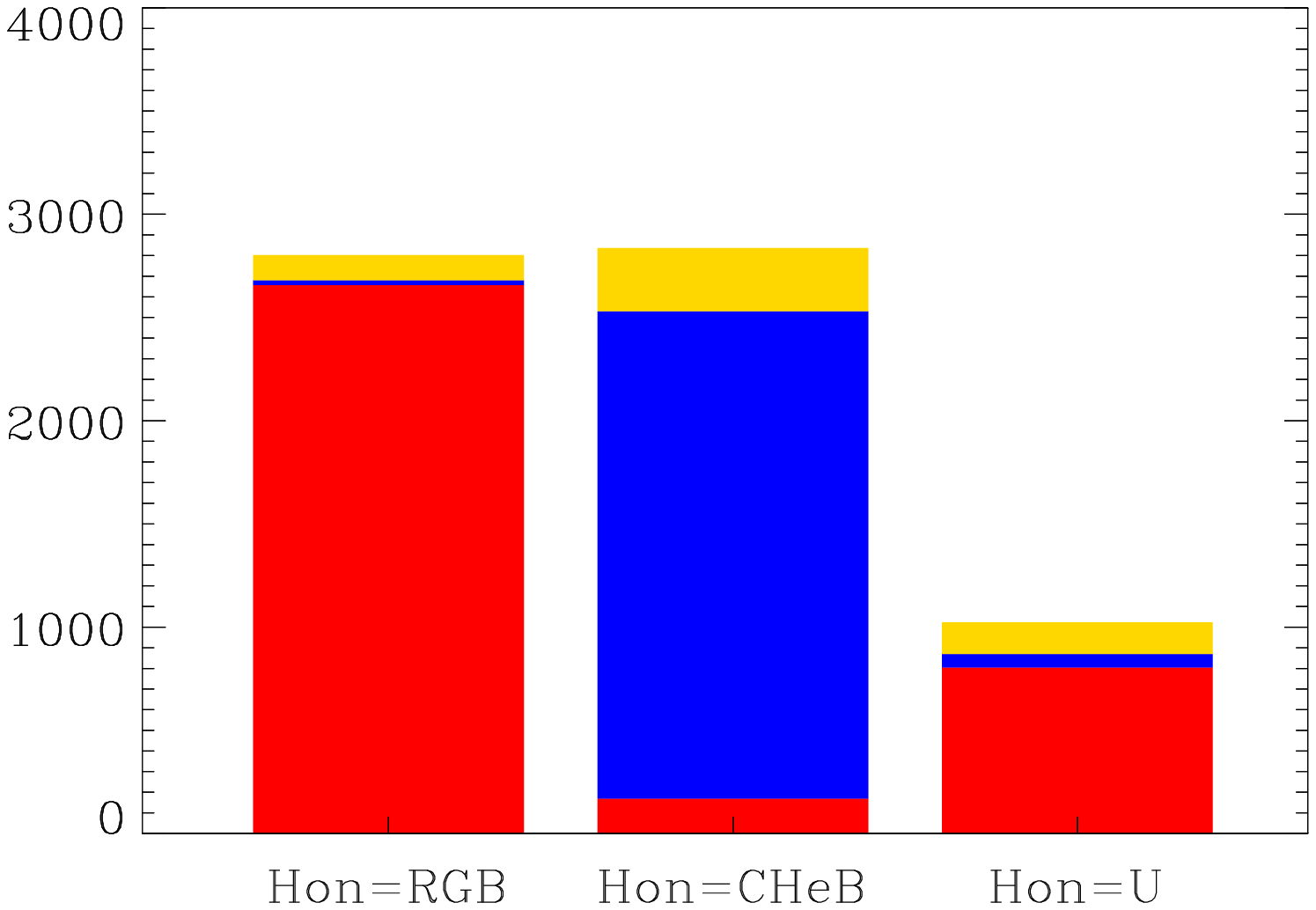}
\caption{A representation of the level of agreement between the \citet{hon2018}  and the consensus classifications.
\citet{hon2018} gives the data in effectively 3 categories, RGB \& CHeB plus some stars are not classified.
Each bar in the chart represents the stars given a particular Hon classification. The colours in the bars indicate how the stars are classified by the consensus method.
The notation used is that red is for RGB stars, blue is for CHeB stars  and yellow is unclassified.}
\label{DA-bars}
\end{figure}

We use the overlap with the `consensus' classifications that are the main topic of this paper, to illustrate these points.
The results of the comparison are shown in Figure~\ref{DA-bars}
The reasons for the unknown classifications in Hon  is that the stars were not considered by the authors,
and we do not consider this further here.

Finally, it should be noted that there may be mis-classifications in the consensus method result but the numbers are expected to be small and so we consider disagreement to be an indication of possible contamination.

\section{Discussion and Conclusions}
\label{summary}
Many important astrophysical problems concerning red-giant stars are greatly helped by knowledge of the evolutionary state of the stars. That is to say, an answer to the question of whether their Helium cores are inert or support Helium fusion.
The addition of asteroseismology to the more usual spectroscopic tool box has enabled this  advance.
We have used  the results of four very different approaches to the determination of the evolutionary state of any given red giant star to produce a consensus value.
It is seen that using a consensus approach improves the reliability of the evolutionary state classification by utilising the diversity of approach to the problem.
The data set presented here is the APOKASC set of \all\ stars.

The result of this analysis is a set of 6197 red giant stars with robustly determined evolutionary states based on evaluating the consensus from four different seismic methods. Because these evolutionary states are based on probes of the stellar interior, they are fundamental and provide a high-fidelity sample of stars that can be used to test other methods and, in particular to calibrate spectroscopic methods as discussed in Section~\ref{spectroscopic}.
This information is especially vital in the regions of the HRD where the \rc\  and \rgb\ overlap.
It should be noted that the spectroscopic method  has the advantage that it can be applied to stars with little or no seismic data.

Additionally, many of these classifications have been used to train recent methods which use machine learning enabling classifications on stars with shorter time series data from mission like K2, TESS, and PLATO (\citet{hon2018}, and \Kusz).

We have noted earlier that in some cases, the seismic classification is difficult.
Here, we mention just two further situations  where the difficulty is due to an
astronomically interesting phenomenon.
One situation is where the strength of the dipole modes is unusually low.
These stars have been widely studied since they were first reported by \citet{mosser2012} and \citet{garcia2014}.
\citet{fuller2015} suggested that a strong fossil magnetic field in the stellar core provides an explanation for the phenomenon \citep[see also][]{stello2016}.
This explanation was questioned by \citet{mosser2017}.

\begin{figure*}
\includegraphics[width=0.95\textwidth]{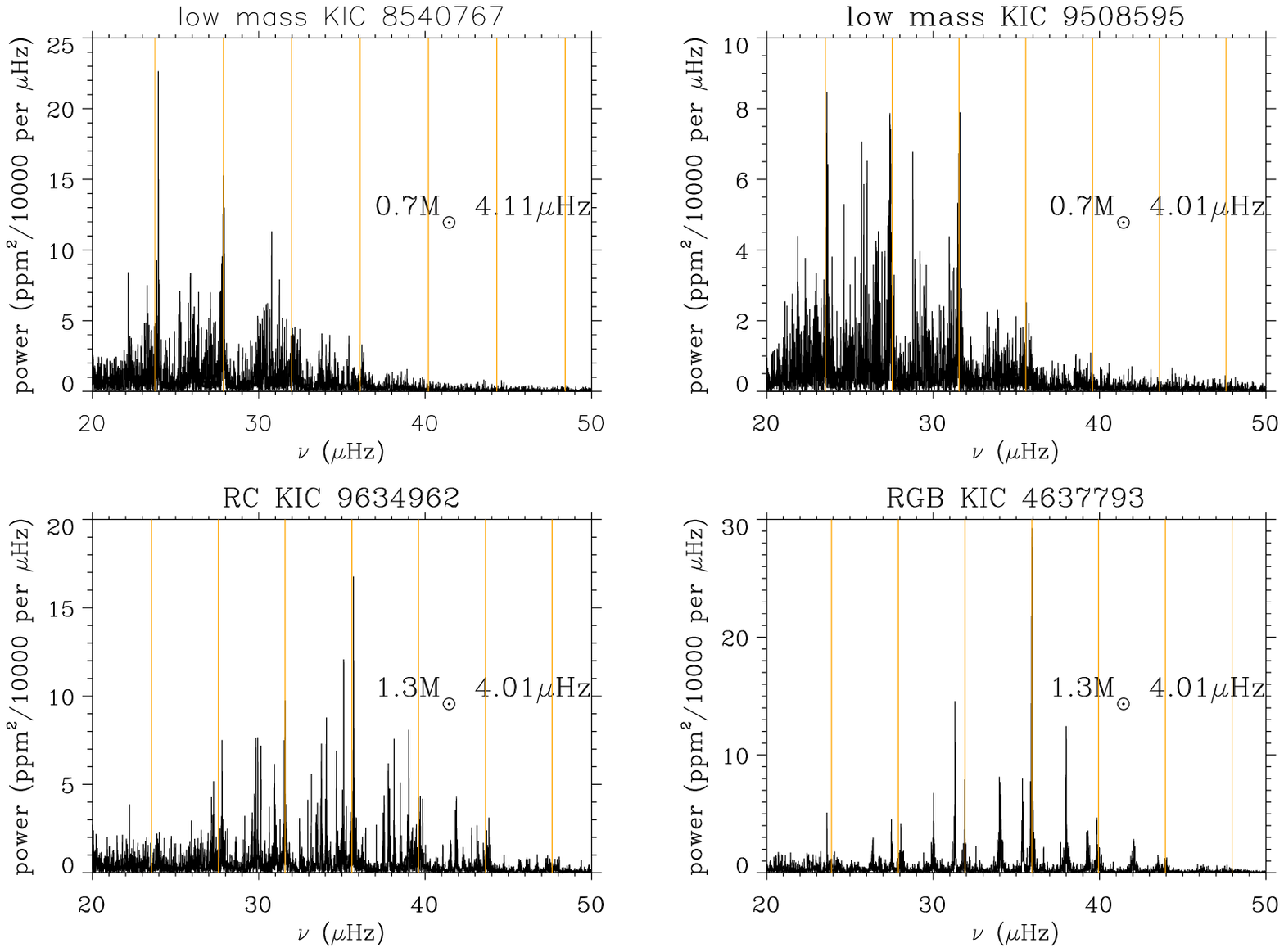}
\caption{Spectra of four stars with differing characteristics. In the upper row are stars with scaling-law masses of about 0.7\Ms.
In the lower row are an RC star and an RGB star with  similar \DeltaNu\ and both with scaling-law masses of about 1.3\Ms.
The numbers in the panels are first the scaling-law mass and then the \DeltaNu\ value.
The vertical yellow lines are indicative of the position of the $\ell=0$ modes. }.
\label{low-mass}
\end{figure*}

Another situation where difficulty may be expected is for stars for which the scaling law masses are below what would be expected in the solar neighbourhood.
Observationally, these stars have masses around 0.7 \Ms.
In the acoustic spectra of these stars, it can be difficult to identify
the quadrupole ($\ell=2$) modes and there is a lack of clear structure in the dipole ($\ell=1$) modes.
A range of typical  spectra are shown in Figure~\ref{low-mass}.
For comparison purposes we show spectra of two low mass stars together with spectra of higher mass stars.
In all cases the \DeltaNu\ values are similar.
The lower row of the figure shows  typical spectra of a \rc\ star (left-hand side) and a first ascent red giant (right-hand side).
Both these  stars have roughly the same scaling-law mass (1.3\Ms ) and \DeltaNu\ (\uHz{4.01}).
On the upper row of the figure are the somewhat unusual spectra of two very low mass stars.
The scaling-law masses of both these is about 0.7\Ms.
These masses are only approximate and some adjustment is required to the scaling-law values. 
However, the morphology of their spectra does suggest that the stars are in the red clump and that the adjustment to the mass is small and positive \citep{sharma2016}.

The right-hand  star has a \DeltaNu\ of \uHz{4.01}.
The  \DeltaNu\ value for the upper left-hand star is slightly larger at \uHz{4.11}.
It is obvious that the spectra of the two very low mass stars are shifted to lower frequencies compared with the 1.3\Ms\ stars.
This is expected.
It is probably significant  that the very low-mass stars have relatively low metallicities at $-$0.67 (RH) and $-$0.15 (LH) as compared with the RGB star with $-$0.06 and the RC star with $+$0.27.
They are also hotter by a few hundred degrees.
There is a suggestions that the lack of a clear mixed mode pattern in the low-mass stars is consistent with the p and the g cavities being very strongly coupled \citep{montalban2013}.
The very low masses, if interpreted simply, would require that the stars are very old.
Perhaps a better option is that they have suffered considerable mass loss in their lives, maybe through earlier binary interaction. These stars offer a good opportunity to study the evolution of the structure of the evanescent zone between the p and g cavities of the stars.

\begin{figure*}
\includegraphics[width=0.95\textwidth]{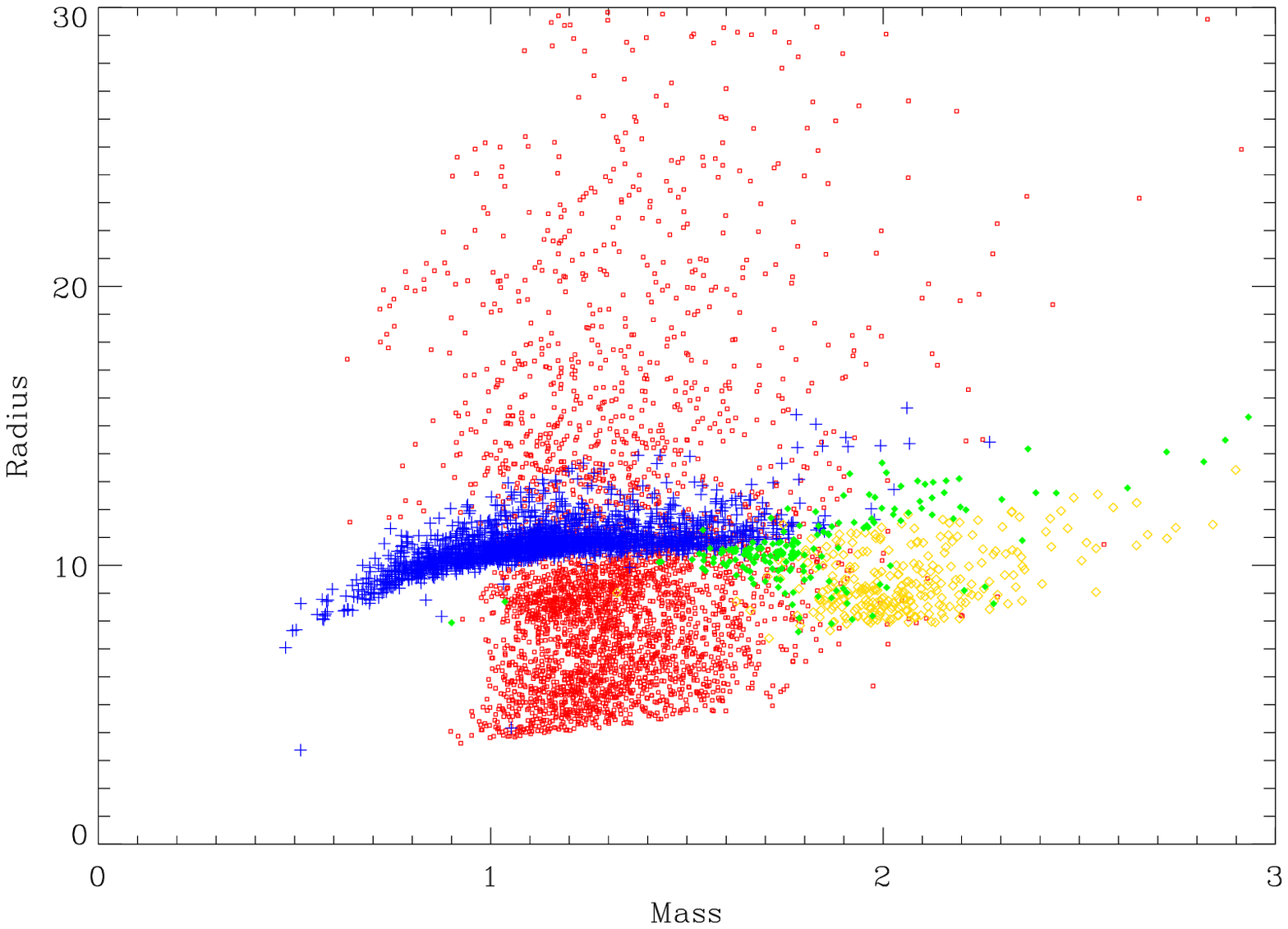}
\caption{A Radius vs. Mass plot for those stars with consensus evolutionary states.
On both axes the values are given in solar units.
The \rgb\ stars are shown as red open  squares.
The stars that are certain \rc\ are shown as blue crosses.
In yellow diamonds, are the stars that are  clear \scl\ and finally, in filled green diamonds are the stars  in the category where there is uncertainty between \rc\ and \scl .
The typical uncertainties on the plotted values are 10\% in mass and 4\% in radius.
} 
\label{rm}
\end{figure*}

Finally, we show the mass and radius data for the stars with consensus evolutionary states in Figure~\ref{rm}.
The \rgb\ stars are visible over a wide range of radii as they evolve up the red-giant branch. 
On the other hand, the \rc\ and \scl\ stars have a much smaller range of radii.
The region occupied by the stars where it was not possible to make a clear decision between \rc\ and \scl\ stars is shown in green. Most of these stars do lie at the interface between these stars with a tendency towards them being \scl\ because of their mass and/or radius.
We also note that there is a hint of a bifurcation in the locations of the clear \scl\ stars.

\section*{Acknowledgments}
AM acknowledges support from the ERC Consolidator Grant funding scheme (project ASTEROCHRONOMETRY, G.A. n. 772293).
AMS is partially supported by grants ESP2017-82674-R (Spanish Government) and 2017-SGR-1131 (Generalitat de Catalunya).
DS is the recipient of an Australian Research Council Future Fellowship (project number FT1400147).
Support for JT was provided by NASA through the NASA Hubble Fellowship grant \#51424 awarded by the Space Telescope Science Institute, which is operated by the Association of Universities for Research in Astronomy, Inc., for NASA, under contract NAS5-26555.
MV acknowledges funding by FCT - Fundação para a Ciência e a Tecnologia  through national funds and by FEDER through COMPETE2020 - Programa Operacional Competitividade e Internacionalização in the context of the grants: PTDC/FIS-AST/30389/2017 \& POCI-01-0145-FEDER-030389 and UID/FIS/04434/2013 \& POCI-01-0145-FEDER-007672.
SH has received funding from the European Research Council under the European Community's Seventh Framework Programme
(FP7/2007-2013) / ERC grant agreement no 338251 (StellarAges).
YE  acknowledges the support of the UK Science and Technology Facilities Council (STFC).
Funding for the Stellar Astrophysics Centre (SAC) is provided by The Danish National Research Foundation (Grant agreement no.: DNRF106).

Funding for the Sloan Digital Sky Survey IV has been provided by the Alfred P. Sloan Foundation, the U.S. Department of Energy Office of Science, and the Participating Institutions. SDSS-IV acknowledges
support and resources from the Center for High-Performance Computing at
the University of Utah. The SDSS web site is www.sdss.org.

SDSS-IV is managed by the Astrophysical Research Consortium for the 
Participating Institutions of the SDSS Collaboration including the 
Brazilian Participation Group, the Carnegie Institution for Science, 
Carnegie Mellon University, the Chilean Participation Group, the French Participation Group, Harvard-Smithsonian Center for Astrophysics, 
Instituto de Astrof\'isica de Canarias, The Johns Hopkins University, Kavli Institute for the Physics and Mathematics of the Universe (IPMU) / 
University of Tokyo, the Korean Participation Group, Lawrence Berkeley National Laboratory, 
Leibniz Institut f\"ur Astrophysik Potsdam (AIP),  
Max-Planck-Institut f\"ur Astronomie (MPIA Heidelberg), 
Max-Planck-Institut f\"ur Astrophysik (MPA Garching), 
Max-Planck-Institut f\"ur Extraterrestrische Physik (MPE), 
National Astronomical Observatories of China, New Mexico State University, 
New York University, University of Notre Dame, 
Observat\'ario Nacional / MCTI, The Ohio State University, 
Pennsylvania State University, Shanghai Astronomical Observatory, 
United Kingdom Participation Group,
Universidad Nacional Aut\'onoma de M\'exico, University of Arizona, 
University of Colorado Boulder, University of Oxford, University of Portsmouth, 
University of Utah, University of Virginia, University of Washington, University of Wisconsin, 
Vanderbilt University, and Yale University.

\appendix
\section{Data}
The data are provided in an on-line table.
The consensus evolutionary state comes only from the asteroseismic data.
Table~\ref{results} is an extract from the full table of results.
The first column is the KIC number of the star considered.
The second column is the consensus evolutionary state from the asteroseismic data. 
The next four columns are the values returned by the individual methods.
The final column is the spectroscopic classification.
The values returned by the spectroscopic method have (S) added to the normal designation for RGB and RC.

\begin{table*}
\begin{tabular}{|c|c||c|c|c|c||c|}
\hline
KIC&consensus&evol-1&evol-2&evol-3&evol-4&spectroscopic\\
\hline
1160789&	RC&	RC&	RC&	RC&	RC&	RC(S)\\
1163621&	RC/2CL&	RC&	RC&	2CL&	RC&	RC(S)\\
1296068&	RGB&	RGB&RGB&	RGB&	RGB&	RGB(S)\\
1726211&	RC&	RC&	RGB&	RC&	RC&	RGB(S)\\
1870433&	RGB&	RGB&	RGB&	RGB&	U&	RGB(S)\\
2164874&	RGB/AGB&	RGB&	RC&	AGB&	RGB&	UpperRGB\\
2444062&	2CL&	U&	RC&	2CL&	2CL&	RC(S)\\
2570715&	RC&	RC&	RGB&	RC&	U&	RC(S)\\
\hline
\end{tabular}
\caption{Table of evolutionary states giving the consensus state and the individual ones for a sample of stars. The full table is available on-line.}
\label{results}
\end{table*}

\bibliographystyle{mn2e_new}

\bibliography{references}

\begin{thebibliography}{79}
\expandafter\ifx\csname natexlab\endcsname\relax\def\natexlab#1{#1}\fi

\bibitem[{{Beck} {et~al}\mbox{.}(2011){Beck}, {Bedding}, {Mosser}, {Stello},
  {Garcia}, {Kallinger}, {Hekker}, {Elsworth}, {Frandsen}, {Carrier}, {De
  Ridder}, {Aerts}, {White}, {Huber}, {Dupret}, {Montalb{\'a}n}, {Miglio},
  {Noels}, {Chaplin}, {Kjeldsen}, {Christensen-Dalsgaard}, {Gilliland},
  {Brown}, {Kawaler}, {Mathur}, \& {Jenkins}}]{beck2011}
{Beck} P.~G. {et~al.}, 2011, Science, 332, 205

\bibitem[{{Bedding} {et~al}\mbox{.}(2010){Bedding}, {Huber}, {Stello},
  {Elsworth}, {Hekker}, {Kallinger}, {Mathur}, {Mosser}, {Preston}, {Ballot},
  {Barban}, {Broomhall}, {Buzasi}, {Chaplin}, {Garc{\'{\i}}a}, {Gruberbauer},
  {Hale}, {De Ridder}, {Frandsen}, {Borucki}, {Brown}, {Christensen-Dalsgaard},
  {Gilliland}, {Jenkins}, {Kjeldsen}, {Koch}, {Belkacem}, {Bildsten}, {Bruntt},
  {Campante}, {Deheuvels}, {Derekas}, {Dupret}, {Goupil}, {Hatzes}, {Houdek},
  {Ireland}, {Jiang}, {Karoff}, {Kiss}, {Lebreton}, {Miglio}, {Montalb{\'a}n},
  {Noels}, {Roxburgh}, {Sangaralingam}, {Stevens}, {Suran}, {Tarrant}, \&
  {Weiss}}]{bedding2010}
{Bedding} T.~R. {et~al.}, 2010, \apjl, 713, L176

\bibitem[{{Bedding} {et~al}\mbox{.}(2011){Bedding}, {Mosser}, {Huber},
  {Montalb{\'a}n}, {Beck}, {Christensen-Dalsgaard}, {Elsworth},
  {Garc{\'{\i}}a}, {Miglio}, {Stello}, {White}, {De Ridder}, {Hekker}, {Aerts},
  {Barban}, {Belkacem}, {Broomhall}, {Brown}, {Buzasi}, {Carrier}, {Chaplin},
  {di Mauro}, {Dupret}, {Frandsen}, {Gilliland}, {Goupil}, {Jenkins},
  {Kallinger}, {Kawaler}, {Kjeldsen}, {Mathur}, {Noels}, {Aguirre}, \&
  {Ventura}}]{bedding2011}
---, 2011, Nature, 471, 608

\bibitem[{{Blanton} {et~al}\mbox{.}(2017){Blanton}, {Bershady}, {Abolfathi},
  {Albareti}, {Allende Prieto}, {Almeida}, {Alonso-Garc{\'{\i}}a}, {Anders},
  {Anderson}, {Andrews}, \& et~al.}]{blanton2017}
{Blanton} M.~R. {et~al.}, 2017, \aj, 154, 28

\bibitem[{{Borucki} {et~al}\mbox{.}(2010){Borucki}, {Koch}, {Basri}, {Batalha},
  {Brown}, {Caldwell}, {Caldwell}, {Christensen-Dalsgaard}, {Cochran},
  {DeVore}, {Dunham}, {Dupree}, {Gautier}, {Geary}, {Gilliland}, {Gould},
  {Howell}, {Jenkins}, {Kondo}, {Latham}, {Marcy}, {Meibom}, {Kjeldsen},
  {Lissauer}, {Monet}, {Morrison}, {Sasselov}, {Tarter}, {Boss}, {Brownlee},
  {Owen}, {Buzasi}, {Charbonneau}, {Doyle}, {Fortney}, {Ford}, {Holman},
  {Seager}, {Steffen}, {Welsh}, {Rowe}, {Anderson}, {Buchhave}, {Ciardi},
  {Walkowicz}, {Sherry}, {Horch}, {Isaacson}, {Everett}, {Fischer}, {Torres},
  {Johnson}, {Endl}, {MacQueen}, {Bryson}, {Dotson}, {Haas}, {Kolodziejczak},
  {Van Cleve}, {Chandrasekaran}, {Twicken}, {Quintana}, {Clarke}, {Allen},
  {Li}, {Wu}, {Tenenbaum}, {Verner}, {Bruhweiler}, {Barnes}, \&
  {Prsa}}]{borucki2010}
{Borucki} W.~J. {et~al.}, 2010, Science, 327, 977

\bibitem[{{Bovy} {et~al}\mbox{.}(2014){Bovy}, {Nidever}, {Rix}, {Girardi},
  {Zasowski}, {Chojnowski}, {Holtzman}, {Epstein}, {Frinchaboy}, {Hayden},
  {Rodrigues}, {Majewski}, {Johnson}, {Pinsonneault}, {Stello}, {Allende
  Prieto}, {Andrews}, {Basu}, {Beers}, {Bizyaev}, {Burton}, {Chaplin}, {Cunha},
  {Elsworth}, {Garc{\'{\i}}a}, {Garc{\'{\i}}a-Her{\'n}andez}, {Garc{\'{\i}}a
  P{\'e}rez}, {Hearty}, {Hekker}, {Kallinger}, {Kinemuchi}, {Koesterke},
  {M{\'e}sz{\'a}ros}, {Mosser}, {O'Connell}, {Oravetz}, {Pan}, {Robin},
  {Schiavon}, {Schneider}, {Schultheis}, {Serenelli}, {Shetrone}, {Silva
  Aguirre}, {Simmons}, {Skrutskie}, {Smith}, {Stassun}, {Weinberg}, {Wilson},
  \& {Zamora}}]{bovy2014}
{Bovy} J. {et~al.}, 2014, \apj, 790, 127

\bibitem[{{Buzzoni} {et~al}\mbox{.}(1983){Buzzoni}, {Pecci}, {Buonanno}, \&
  {Corsi}}]{buzzoni1983}
{Buzzoni} A., {Pecci} F.~F., {Buonanno} R., {Corsi} C.~E., 1983, \aap, 128, 94

\bibitem[{{Casagrande} {et~al}\mbox{.}(2016){Casagrande}, {Silva Aguirre},
  {Schlesinger}, {Stello}, {Huber}, {Serenelli}, {Sch{\"o}nrich}, {Cassisi},
  {Pietrinferni}, {Hodgkin}, {Milone}, {Feltzing}, \&
  {Asplund}}]{casagrande2016}
{Casagrande} L. {et~al.}, 2016, \mnras, 455, 987

\bibitem[{{Christensen-Dalsgaard}(2014)}]{jcd2014book}
{Christensen-Dalsgaard} J., 2014, in Asteroseismology, {Pall{\'e}} P.~L.,
  {Esteban} C., eds., Cambridge University Press, p. 194

\bibitem[{{Colman} {et~al}\mbox{.}(2017){Colman}, {Huber}, {Bedding},
  {Kuszlewicz}, {Yu}, {Beck}, {Elsworth}, {Garc{\'{\i}}a}, {Kawaler}, {Mathur},
  {Stello}, \& {White}}]{colman2017}
{Colman} I.~L. {et~al.}, 2017, \mnras, 469, 3802

\bibitem[{{Constantino} {et~al}\mbox{.}(2016){Constantino}, {Campbell},
  {Lattanzio}, \& {van Duijneveldt}}]{constantino2015}
{Constantino} T., {Campbell} S.~W., {Lattanzio} J.~C., {van Duijneveldt} A.,
  2016, \mnras, 456, 3866

\bibitem[{{Davies} {et~al}\mbox{.}(2017){Davies}, {Lund}, {Miglio}, {Elsworth},
  {Kuszlewicz}, {North}, {Rendle}, {Chaplin}, {Rodrigues}, {Campante},
  {Girardi}, {Hale}, {Hall}, {Jones}, {Kawaler}, {Roxburgh}, \&
  {Schofield}}]{davies2017}
{Davies} G.~R. {et~al.}, 2017, \aap, 598, L4

\bibitem[{{De Ridder} {et~al}\mbox{.}(2009){De Ridder}, {Barban}, {Baudin},
  {Carrier}, {Hatzes}, {Hekker}, {Kallinger}, {Weiss}, {Baglin}, {Auvergne},
  {Samadi}, {Barge}, \& {Deleuil}}]{deridder2009}
{De Ridder} J. {et~al.}, 2009, \nat, 459, 398

\bibitem[{{Dupret} {et~al}\mbox{.}(2009){Dupret}, {Belkacem}, {Samadi},
  {Montalban}, {Moreira}, {Miglio}, {Godart}, {Ventura}, {Ludwig},
  {Grigahc{\`e}ne}, {Goupil}, {Noels}, \& {Caffau}}]{dupret2009}
{Dupret} M.-A. {et~al.}, 2009, \aap, 506, 57

\bibitem[{{Eisenstein} {et~al}\mbox{.}(2011){Eisenstein}, {Weinberg}, {Agol},
  {Aihara}, {Allende Prieto}, {Anderson}, {Arns}, {Aubourg}, {Bailey},
  {Balbinot}, \& et~al.}]{eisenstein2011}
{Eisenstein} D.~J. {et~al.}, 2011, \aj, 142, 72

\bibitem[{{Elsworth} {et~al}\mbox{.}(2017){Elsworth}, {Hekker}, {Basu}, \&
  {Davies}}]{elsworth2017evolYE}
{Elsworth} Y., {Hekker} S., {Basu} S., {Davies} G.~R., 2017, \mnras, 466, 3344

\bibitem[{{Fuller} {et~al}\mbox{.}(2015){Fuller}, {Cantiello}, {Stello},
  {Garcia}, \& {Bildsten}}]{fuller2015}
{Fuller} J., {Cantiello} M., {Stello} D., {Garcia} R.~A., {Bildsten} L., 2015,
  Science, 350, 423

\bibitem[{{Gaia Collaboration} {et~al}\mbox{.}(2018){Gaia Collaboration},
  {Brown}, {Vallenari}, {Prusti}, {de Bruijne}, {Babusiaux}, {Bailer-Jones},
  {Biermann}, {Evans}, {Eyer}, \& et~al.}]{gaia2018}
{Gaia Collaboration} {et~al.}, 2018, \aap, 616, A1

\bibitem[{{Gaia Collaboration} {et~al}\mbox{.}(2016){Gaia Collaboration},
  {Prusti}, {de Bruijne}, {Brown}, {Vallenari}, {Babusiaux}, {Bailer-Jones},
  {Bastian}, {Biermann}, {Evans}, \& et~al.}]{gaia2016}
---, 2016, \aap, 595, A1

\bibitem[{{Garc{\'{\i}}a} {et~al}\mbox{.}(2014){Garc{\'{\i}}a}, {P{\'e}rez
  Hern{\'a}ndez}, {Benomar}, {Silva Aguirre}, {Ballot}, {Davies}, {Do{\u g}an},
  {Stello}, {Christensen-Dalsgaard}, {Houdek}, {Ligni{\`e}res}, {Mathur},
  {Takata}, {Ceillier}, {Chaplin}, {Mathis}, {Mosser}, {Ouazzani},
  {Pinsonneault}, {Reese}, {R{\'e}gulo}, {Salabert}, {Thompson}, {van Saders},
  {Neiner}, \& {De Ridder}}]{garcia2014}
{Garc{\'{\i}}a} R.~A. {et~al.}, 2014, \aap, 563, A84

\bibitem[{{Garc{\'{\i}}a P{\'e}rez} {et~al}\mbox{.}(2016){Garc{\'{\i}}a
  P{\'e}rez}, {Allende Prieto}, {Holtzman}, {Shetrone}, {M{\'e}sz{\'a}ros},
  {Bizyaev}, {Carrera}, {Cunha}, {Garc{\'{\i}}a-Hern{\'a}ndez}, {Johnson},
  {Majewski}, {Nidever}, {Schiavon}, {Shane}, {Smith}, {Sobeck}, {Troup},
  {Zamora}, {Weinberg}, {Bovy}, {Eisenstein}, {Feuillet}, {Frinchaboy},
  {Hayden}, {Hearty}, {Nguyen}, {O'Connell}, {Pinsonneault}, {Wilson}, \&
  {Zasowski}}]{garciaPerez2016}
{Garc{\'{\i}}a P{\'e}rez} A.~E. {et~al.}, 2016, \aj, 151, 144

\bibitem[{{Gaulme} {et~al}\mbox{.}(2014){Gaulme}, {Jackiewicz}, {Appourchaux},
  \& {Mosser}}]{gaulme2014}
{Gaulme} P., {Jackiewicz} J., {Appourchaux} T., {Mosser} B., 2014, \apj, 785, 5

\bibitem[{{Girardi}(1999)}]{girardi1999}
{Girardi} L., 1999, \mnras, 308, 818

\bibitem[{{Grosjean} {et~al}\mbox{.}(2014){Grosjean}, {Dupret}, {Belkacem},
  {Montalban}, {Samadi}, \& {Mosser}}]{grosjean2014}
{Grosjean} M., {Dupret} M.-A., {Belkacem} K., {Montalban} J., {Samadi} R.,
  {Mosser} B., 2014, \aap, 572, A11

\bibitem[{{Gunn} {et~al}\mbox{.}(2006){Gunn}, {Siegmund}, {Mannery}, {Owen},
  {Hull}, {Leger}, {Carey}, {Knapp}, {York}, {Boroski}, {Kent}, {Lupton},
  {Rockosi}, {Evans}, {Waddell}, {Anderson}, {Annis}, {Barentine}, {Bartoszek},
  {Bastian}, {Bracker}, {Brewington}, {Briegel}, {Brinkmann}, {Brown}, {Carr},
  {Czarapata}, {Drennan}, {Dombeck}, {Federwitz}, {Gillespie}, {Gonzales},
  {Hansen}, {Harvanek}, {Hayes}, {Jordan}, {Kinney}, {Klaene}, {Kleinman},
  {Kron}, {Kresinski}, {Lee}, {Limmongkol}, {Lindenmeyer}, {Long}, {Loomis},
  {McGehee}, {Mantsch}, {Neilsen}, {Neswold}, {Newman}, {Nitta}, {Peoples},
  {Pier}, {Prieto}, {Prosapio}, {Rivetta}, {Schneider}, {Snedden}, \&
  {Wang}}]{gunn2006}
{Gunn} J.~E. {et~al.}, 2006, \aj, 131, 2332

\bibitem[{{Hasselquist} {et~al}\mbox{.}(2019){Hasselquist}, {Holtzman},
  {Shetrone}, {Tayar}, {Weinberg}, {Feuillet}, {Cunha}, {Pinsonneault},
  {Johnson}, {Bird}, {Beers}, {Schiavon}, {Minchev}, {Fern{\'a}ndez-Trincado},
  {Garc{\'{\i}}a-Hern{\'a}ndez}, {Nitschelm}, \& {Zamora}}]{hasselquist2019}
{Hasselquist} S. {et~al.}, 2019, \apj, 871, 181

\bibitem[{{Hawkins} {et~al}\mbox{.}(2016){Hawkins}, {Masseron}, {Jofr{\'e}},
  {Gilmore}, {Elsworth}, \& {Hekker}}]{hawkins2016}
{Hawkins} K., {Masseron} T., {Jofr{\'e}} P., {Gilmore} G., {Elsworth} Y.,
  {Hekker} S., 2016, \aap, 594, A43

\bibitem[{{Hekker}(2019)}]{hekker2019}
{Hekker} S., 2019, arXiv e-prints, arXiv:1907.10457

\bibitem[{{Hekker} \& {Christensen-Dalsgaard}(2017)}]{hekkerJCD2017}
{Hekker} S., {Christensen-Dalsgaard} J., 2017, \aapr, 25, 1

\bibitem[{{Hekker} {et~al}\mbox{.}(2017){Hekker}, {Elsworth}, {Basu}, \&
  {Bellinger}}]{hekker2017svm}
{Hekker} S., {Elsworth} Y., {Basu} S., {Bellinger} E., 2017, ArXiv e-prints

\bibitem[{{Hekker} \& {Johnson}(2019)}]{hekkerJohnson2019}
{Hekker} S., {Johnson} J.~A., 2019, \mnras, 487, 4343

\bibitem[{{Hekker} {et~al}\mbox{.}(2009){Hekker}, {Kallinger}, {Baudin}, {De
  Ridder}, {Barban}, {Carrier}, {Hatzes}, {Weiss}, \& {Baglin}}]{hekker2009}
{Hekker} S. {et~al.}, 2009, \aap, 506, 465

\bibitem[{{Holtzman} {et~al}\mbox{.}(2018){Holtzman}, {Hasselquist},
  {Shetrone}, {Cunha}, {Allende Prieto}, {Anguiano}, {Bizyaev}, {Bovy},
  {Casey}, {Edvardsson}, {Johnson}, {J{\"o}nsson}, {Meszaros}, {Smith},
  {Sobeck}, {Zamora}, {Chojnowski}, {Fernandez-Trincado}, {Garcia-Hernandez},
  {Majewski}, {Pinsonneault}, {Souto}, {Stringfellow}, {Tayar}, {Troup}, \&
  {Zasowski}}]{holtzman2018}
{Holtzman} J.~A. {et~al.}, 2018, \aj, 156, 125

\bibitem[{{Hon}, {Stello} \& {Yu}(2018){Hon}, {Stello}, \& {Yu}}]{hon2018}
{Hon} M., {Stello} D., {Yu} J., 2018, \mnras, 476, 3233

\bibitem[{{Iben} \& {Rood}(1969)}]{ibenRood1969}
{Iben} I., {Rood} R.~T., 1969, \nat, 223, 933

\bibitem[{{Iben}(1965)}]{iben1965II}
{Iben}, Jr. I., 1965, \apj, 142, 1447

\bibitem[{{Izzard} {et~al}\mbox{.}(2018){Izzard}, {Preece}, {Jofre}, {Halabi},
  {Masseron}, \& {Tout}}]{izzard2018}
{Izzard} R.~G., {Preece} H., {Jofre} P., {Halabi} G.~M., {Masseron} T., {Tout}
  C.~A., 2018, \mnras, 473, 2984

\bibitem[{{Jofr{\'e}} {et~al}\mbox{.}(2016){Jofr{\'e}}, {Jorissen}, {Van Eck},
  {Izzard}, {Masseron}, {Hawkins}, {Gilmore}, {Paladini}, {Escorza},
  {Blanco-Cuaresma}, \& {Manick}}]{jofre2016}
{Jofr{\'e}} P. {et~al.}, 2016, \aap, 595, A60

\bibitem[{{Kallinger} {et~al}\mbox{.}(2012){Kallinger}, {Hekker}, {Mosser}, {De
  Ridder}, {Bedding}, {Elsworth}, {Gruberbauer}, {Guenther}, {Stello}, {Basu},
  {Garc{\'{\i}}a}, {Chaplin}, {Mullally}, {Still}, \&
  {Thompson}}]{kallinger2012}
{Kallinger} T. {et~al.}, 2012, A\&A, 541, A51

\bibitem[{{Kjeldsen} \& {Bedding}(1995)}]{kb1995}
{Kjeldsen} H., {Bedding} T.~R., 1995, \aap, 293

\bibitem[{{Kraft}(1994)}]{kraft1994}
{Kraft} R.~P., 1994, \pasp, 106, 553

\bibitem[{{Majewski} {et~al}\mbox{.}(2017){Majewski}, {Schiavon}, {Frinchaboy},
  {Allende Prieto}, {Barkhouser}, {Bizyaev}, {Blank}, {Brunner}, {Burton},
  {Carrera}, {Chojnowski}, {Cunha}, {Epstein}, {Fitzgerald}, {Garc{\'{\i}}a
  P{\'e}rez}, {Hearty}, {Henderson}, {Holtzman}, {Johnson}, {Lam}, {Lawler},
  {Maseman}, {M{\'e}sz{\'a}ros}, {Nelson}, {Nguyen}, {Nidever}, {Pinsonneault},
  {Shetrone}, {Smee}, {Smith}, {Stolberg}, {Skrutskie}, {Walker}, {Wilson},
  {Zasowski}, {Anders}, {Basu}, {Beland}, {Blanton}, {Bovy}, {Brownstein},
  {Carlberg}, {Chaplin}, {Chiappini}, {Eisenstein}, {Elsworth}, {Feuillet},
  {Fleming}, {Galbraith-Frew}, {Garc{\'{\i}}a}, {Garc{\'{\i}}a-Hern{\'a}ndez},
  {Gillespie}, {Girardi}, {Gunn}, {Hasselquist}, {Hayden}, {Hekker}, {Ivans},
  {Kinemuchi}, {Klaene}, {Mahadevan}, {Mathur}, {Mosser}, {Muna}, {Munn},
  {Nichol}, {O'Connell}, {Parejko}, {Robin}, {Rocha-Pinto}, {Schultheis},
  {Serenelli}, {Shane}, {Silva Aguirre}, {Sobeck}, {Thompson}, {Troup},
  {Weinberg}, \& {Zamora}}]{majewski2017}
{Majewski} S.~R. {et~al.}, 2017, \aj, 154, 94

\bibitem[{{Martig} {et~al}\mbox{.}(2016){Martig}, {Fouesneau}, {Rix}, {Ness},
  {M{\'e}sz{\'a}ros}, {Garc{\'{\i}}a-Hern{\'a}ndez}, {Pinsonneault},
  {Serenelli}, {Silva Aguirre}, \& {Zamora}}]{martig2016}
{Martig} M. {et~al.}, 2016, \mnras, 456, 3655

\bibitem[{{Masseron} {et~al}\mbox{.}(2017){Masseron}, {Lagarde}, {Miglio},
  {Elsworth}, \& {Gilmore}}]{masseron2017a}
{Masseron} T., {Lagarde} N., {Miglio} A., {Elsworth} Y., {Gilmore} G., 2017,
  \mnras, 464, 3021

\bibitem[{{Mathur} {et~al}\mbox{.}(2011){Mathur}, {Hekker}, {Trampedach},
  {Ballot}, {Kallinger}, {Buzasi}, {Garc{\'{\i}}a}, {Huber}, {Jim{\'e}nez},
  {Mosser}, {Bedding}, {Elsworth}, {R{\'e}gulo}, {Stello}, {Chaplin}, {De
  Ridder}, {Hale}, {Kinemuchi}, {Kjeldsen}, {Mullally}, \&
  {Thompson}}]{mathur2011}
{Mathur} S. {et~al.}, 2011, \apj, 741, 119

\bibitem[{{Miglio} {et~al}\mbox{.}(2012){Miglio}, {Brogaard}, {Stello},
  {Chaplin}, {D'Antona}, {Montalb{\'a}n}, {Basu}, {Bressan}, {Grundahl},
  {Pinsonneault}, {Serenelli}, {Elsworth}, {Hekker}, {Kallinger}, {Mosser},
  {Ventura}, {Bonanno}, {Noels}, {Silva Aguirre}, {Szabo}, {Li}, {McCauliff},
  {Middour}, \& {Kjeldsen}}]{miglio2012}
{Miglio} A. {et~al.}, 2012, \mnras, 419, 2077

\bibitem[{{Minniti}(1995)}]{minniti1995}
{Minniti} D., 1995, \aap, 300, 109

\bibitem[{{Montalb{\'a}n} {et~al}\mbox{.}(2013){Montalb{\'a}n}, {Miglio},
  {Noels}, {Dupret}, {Scuflaire}, \& {Ventura}}]{montalban2013}
{Montalb{\'a}n} J., {Miglio} A., {Noels} A., {Dupret} M.-A., {Scuflaire} R.,
  {Ventura} P., 2013, \apj, 766, 118

\bibitem[{{Montalb{\'a}n} {et~al}\mbox{.}(2010){Montalb{\'a}n}, {Miglio},
  {Noels}, {Scuflaire}, \& {Ventura}}]{montalban2010}
{Montalb{\'a}n} J., {Miglio} A., {Noels} A., {Scuflaire} R., {Ventura} P.,
  2010, ApJ, 721, L182

\bibitem[{{Mosser} \& {Appourchaux}(2009)}]{mosser2009}
{Mosser} B., {Appourchaux} T., 2009, \aap, 508, 877

\bibitem[{{Mosser} {et~al}\mbox{.}(2011{\natexlab{a}}){Mosser}, {Barban},
  {Montalb{\'a}n}, {Beck}, {Miglio}, {Belkacem}, {Goupil}, {Hekker}, {De
  Ridder}, {Dupret}, {Elsworth}, {Noels}, {Baudin}, {Michel}, {Samadi},
  {Auvergne}, {Baglin}, \& {Catala}}]{mosser2011mm}
{Mosser} B. {et~al.}, 2011{\natexlab{a}}, A\&A, 532, A86

\bibitem[{{Mosser} {et~al}\mbox{.}(2011{\natexlab{b}}){Mosser}, {Belkacem},
  {Goupil}, {Michel}, {Elsworth}, {Barban}, {Kallinger}, {Hekker}, {De Ridder},
  {Samadi}, {Baudin}, {Pinheiro}, {Auvergne}, {Baglin}, \&
  {Catala}}]{mosser2011up}
---, 2011{\natexlab{b}}, A\&A, 525, L9

\bibitem[{{Mosser} {et~al}\mbox{.}(2017){Mosser}, {Belkacem}, {Pin{\c c}on},
  {Takata}, {Vrard}, {Barban}, {Goupil}, {Kallinger}, \& {Samadi}}]{mosser2017}
---, 2017, \aap, 598, A62

\bibitem[{{Mosser} {et~al}\mbox{.}(2014){Mosser}, {Benomar}, {Belkacem},
  {Goupil}, {Lagarde}, {Michel}, {Lebreton}, {Stello}, {Vrard}, {Barban},
  {Bedding}, {Deheuvels}, {Chaplin}, {De Ridder}, {Elsworth}, {Montalban},
  {Noels}, {Ouazzani}, {Samadi}, {White}, \& {Kjeldsen}}]{mosser2014}
---, 2014, \aap, 572, L5

\bibitem[{{Mosser} {et~al}\mbox{.}(2012{\natexlab{a}}){Mosser}, {Elsworth},
  {Hekker}, {Huber}, {Kallinger}, {Mathur}, {Belkacem}, {Goupil}, {Samadi},
  {Barban}, {Bedding}, {Chaplin}, {Garc{\'{\i}}a}, {Stello}, {De Ridder},
  {Middour}, {Morris}, \& {Quintana}}]{mosser2012}
---, 2012{\natexlab{a}}, \aap, 537, A30

\bibitem[{{Mosser} {et~al}\mbox{.}(2018){Mosser}, {Gehan}, {Belkacem},
  {Samadi}, {Michel}, \& {Goupil}}]{mosser2018}
{Mosser} B., {Gehan} C., {Belkacem} K., {Samadi} R., {Michel} E., {Goupil}
  M.-J., 2018, \aap, 618, A109

\bibitem[{{Mosser} {et~al}\mbox{.}(2012{\natexlab{b}}){Mosser}, {Goupil},
  {Belkacem}, {Michel}, {Stello}, {Marques}, {Elsworth}, {Barban}, {Beck},
  {Bedding}, {De Ridder}, {Garc{\'{\i}}a}, {Hekker}, {Kallinger}, {Samadi},
  {Stumpe}, {Barclay}, \& {Burke}}]{mosser2012core}
{Mosser} B. {et~al.}, 2012{\natexlab{b}}, \aap, 540, A143

\bibitem[{{Mosser} {et~al}\mbox{.}(2015){Mosser}, {Vrard}, {Belkacem},
  {Deheuvels}, \& {Goupil}}]{mosser2015}
{Mosser} B., {Vrard} M., {Belkacem} K., {Deheuvels} S., {Goupil} M.~J., 2015,
  \aap, 584, A50

\bibitem[{{Ness} {et~al}\mbox{.}(2016){Ness}, {Hogg}, {Rix}, {Martig},
  {Pinsonneault}, \& {Ho}}]{ness2016}
{Ness} M., {Hogg} D.~W., {Rix} H.-W., {Martig} M., {Pinsonneault} M.~H., {Ho}
  A.~Y.~Q., 2016, \apj, 823, 114

\bibitem[{{Nidever} {et~al}\mbox{.}(2014){Nidever}, {Bovy}, {Bird}, {Andrews},
  {Hayden}, {Holtzman}, {Majewski}, {Smith}, {Robin}, {Garc{\'{\i}}a
  P{\'e}rez}, {Cunha}, {Allende Prieto}, {Zasowski}, {Schiavon}, {Johnson},
  {Weinberg}, {Feuillet}, {Schneider}, {Shetrone}, {Sobeck},
  {Garc{\'{\i}}a-Hern{\'a}ndez}, {Zamora}, {Rix}, {Beers}, {Wilson},
  {O'Connell}, {Minchev}, {Chiappini}, {Anders}, {Bizyaev}, {Brewington},
  {Ebelke}, {Frinchaboy}, {Ge}, {Kinemuchi}, {Malanushenko}, {Malanushenko},
  {Marchante}, {M{\'e}sz{\'a}ros}, {Oravetz}, {Pan}, {Simmons}, \&
  {Skrutskie}}]{nidever2014}
{Nidever} D.~L. {et~al.}, 2014, \apj, 796, 38

\bibitem[{{Nidever} {et~al}\mbox{.}(2015){Nidever}, {Holtzman}, {Allende
  Prieto}, {Beland}, {Bender}, {Bizyaev}, {Burton}, {Desphande}, {Fleming},
  {Garc{\'{\i}}a P{\'e}rez}, {Hearty}, {Majewski}, {M{\'e}sz{\'a}ros}, {Muna},
  {Nguyen}, {Schiavon}, {Shetrone}, {Skrutskie}, {Sobeck}, \&
  {Wilson}}]{nidever2015}
---, 2015, \aj, 150, 173

\bibitem[{{Perryman} {et~al}\mbox{.}(1997){Perryman}, {Lindegren},
  {Kovalevsky}, {Hoeg}, {Bastian}, {Bernacca}, {Cr{\'e}z{\'e}}, {Donati},
  {Grenon}, {Grewing}, {van Leeuwen}, {van der Marel}, {Mignard}, {Murray}, {Le
  Poole}, {Schrijver}, {Turon}, {Arenou}, {Froeschl{\'e}}, \&
  {Petersen}}]{perryman1997}
{Perryman} M.~A.~C. {et~al.}, 1997, \aap, 323, L49

\bibitem[{{Pinsonneault} {et~al}\mbox{.}(2018){Pinsonneault}, {Elsworth},
  {Silva Aguirre}, {Chaplin}, {Garcia}, {Hekker}, {Holtzman}, {Huber},
  {Johnson}, {Kallinger}, {Mosser}, {Mathur}, {Serenelli}, {Shetrone},
  {Stello}, {Tayar}, {Zinn}, \& {APOGEE Team}}]{pinsonneault2018}
{Pinsonneault} M. {et~al.}, 2018, in American Astronomical Society Meeting
  Abstracts, Vol. 231, American Astronomical Society Meeting Abstracts \#231,
  p. \#450.13

\bibitem[{{Pinsonneault} {et~al}\mbox{.}(2014){Pinsonneault}, {Elsworth},
  {Epstein}, {Hekker}, {M{\'e}sz{\'a}ros}, {Chaplin}, {Johnson},
  {Garc{\'{\i}}a}, {Holtzman}, {Mathur}, {Garc{\'{\i}}a P{\'e}rez}, {Silva
  Aguirre}, {Girardi}, {Basu}, {Shetrone}, {Stello}, {Allende Prieto}, {An},
  {Beck}, {Beers}, {Bizyaev}, {Bloemen}, {Bovy}, {Cunha}, {De Ridder},
  {Frinchaboy}, {Garc{\'{\i}}a-Hern{\'a}ndez}, {Gilliland}, {Harding},
  {Hearty}, {Huber}, {Ivans}, {Kallinger}, {Majewski}, {Metcalfe}, {Miglio},
  {Mosser}, {Muna}, {Nidever}, {Schneider}, {Serenelli}, {Smith}, {Tayar},
  {Zamora}, \& {Zasowski}}]{pinsonneault2014}
{Pinsonneault} M.~H. {et~al.}, 2014, \apjs, 215, 19

\bibitem[{{Renzini} \& {Ritossa}(1994)}]{renzini1994}
{Renzini} A., {Ritossa} C., 1994, \apj, 433, 293

\bibitem[{{Ricker} {et~al}\mbox{.}(2014){Ricker}, {Vanderspek}, {Latham}, \&
  {Winn}}]{ricker2014}
{Ricker} G.~R., {Vanderspek} R.~K., {Latham} D.~W., {Winn} J.~N., 2014, in
  American Astronomical Society Meeting Abstracts, Vol. 224, American
  Astronomical Society Meeting Abstracts \#224, p. 113.02

\bibitem[{{Rodrigues} {et~al}\mbox{.}(2017){Rodrigues}, {Bossini}, {Miglio},
  {Girardi}, {Montalb{\'a}n}, {Noels}, {Trabucchi}, {Coelho}, \&
  {Marigo}}]{rodrigues2017}
{Rodrigues} T.~S. {et~al.}, 2017, \mnras, 467, 1433

\bibitem[{{Salaris} {et~al}\mbox{.}(2015){Salaris}, {Pietrinferni},
  {Piersimoni}, \& {Cassisi}}]{salaris2015Dredge}
{Salaris} M., {Pietrinferni} A., {Piersimoni} A.~M., {Cassisi} S., 2015, \aap,
  583, A87

\bibitem[{{Seidel}, {Demarque} \& {Weinberg}(1987){Seidel}, {Demarque}, \&
  {Weinberg}}]{seidel1987}
{Seidel} E., {Demarque} P., {Weinberg} D., 1987, \apjs, 63, 917

\bibitem[{{Sharma} {et~al}\mbox{.}(2016){Sharma}, {Stello}, {Bland-Hawthorn},
  {Huber}, \& {Bedding}}]{sharma2016}
{Sharma} S., {Stello} D., {Bland-Hawthorn} J., {Huber} D., {Bedding} T.~R.,
  2016, \apj, 822, 15

\bibitem[{{Shetrone} {et~al}\mbox{.}(2019){Shetrone}, {Tayar}, {Johnson},
  {Somers}, {Pinsonneault}, {Holtzman}, {Hasselquist}, {Masseron},
  {M{\'e}sz{\'a}ros}, {J{\"o}nsson}, {Hawkins}, {Sobeck}, {Zamora}, \&
  {Garc{\'{\i}}a-Hern{\'a}ndez}}]{shetrone2019}
{Shetrone} M. {et~al.}, 2019, \apj, 872, 137

\bibitem[{{Stello} {et~al}\mbox{.}(2016){Stello}, {Cantiello}, {Fuller},
  {Huber}, {Garc{\'{\i}}a}, {Bedding}, {Bildsten}, \& {Silva
  Aguirre}}]{stello2016}
{Stello} D., {Cantiello} M., {Fuller} J., {Huber} D., {Garc{\'{\i}}a} R.~A.,
  {Bedding} T.~R., {Bildsten} L., {Silva Aguirre} V., 2016, \nat, 529, 364

\bibitem[{{Stello} {et~al}\mbox{.}(2013){Stello}, {Huber}, {Bedding},
  {Benomar}, {Bildsten}, {Elsworth}, {Gilliland}, {Mosser}, {Paxton}, \&
  {White}}]{stello2013}
{Stello} D. {et~al.}, 2013, ApJ, 765, L41

\bibitem[{{Tautvai{\v s}ien{\.e}} {et~al}\mbox{.}(2015){Tautvai{\v s}ien{\.e}},
  {Drazdauskas}, {Mikolaitis}, {Barisevi{\v c}ius}, {Puzeras}, {Stonkut{\.e}},
  {Chorniy}, {Magrini}, {Romano}, {Smiljanic}, {Bragaglia}, {Carraro}, {Friel},
  {Morel}, {Pancino}, {Donati}, {Jim{\'e}nez-Esteban}, {Gilmore}, {Randich},
  {Jeffries}, {Vallenari}, {Bensby}, {Flaccomio}, {Recio-Blanco}, {Costado},
  {Hill}, {Jofr{\'e}}, {Lardo}, {de Laverny}, {Masseron}, {Moribelli}, {Sousa},
  \& {Zaggia}}]{tautvaisiene2015}
{Tautvai{\v s}ien{\.e}} G. {et~al.}, 2015, \aap, 573, A55

\bibitem[{{Tiede}, {Frogel} \& {Terndrup}(1995){Tiede}, {Frogel}, \&
  {Terndrup}}]{tiede1995}
{Tiede} G.~P., {Frogel} J.~A., {Terndrup} D.~M., 1995, \aj, 110, 2788

\bibitem[{{Vrard}, {Mosser} \& {Samadi}(2016){Vrard}, {Mosser}, \&
  {Samadi}}]{vrard2016}
{Vrard} M., {Mosser} B., {Samadi} R., 2016, \aap, 588, A87

\bibitem[{{Wilson} {et~al}\mbox{.}(2019){Wilson}, {Hearty}, {Skrutskie},
  {Majewski}, {Holtzman}, {Eisenstein}, {Gunn}, {Blank}, {Henderson}, {Smee},
  {Nelson}, {Nidever}, {Arns}, {Barkhouser}, {Barr}, {Beland}, {Bershady},
  {Blanton}, {Brunner}, {Burton}, {Carey}, {Carr}, {Colque}, {Crane}, {Damke},
  {Davidson}, {Dean}, {Di Mille}, {Don}, {Ebelke}, {Evans}, {Fitzgerald},
  {Gillespie}, {Hall}, {Harding}, {Harding}, {Hammond}, {Hancock}, {Harrison},
  {Hope}, {Horne}, {Karakla}, {Lam}, {Leger}, {MacDonald}, {Maseman},
  {Matsunari}, {Melton}, {Mitcheltree}, {O'Brien}, {O'Connell}, {Patten},
  {Richardson}, {Rieke}, {Rieke}, {Roman-Lopes}, {Schiavon}, {Sobeck},
  {Stolberg}, {Stoll}, {Tembe}, {Trujillo}, {Uomoto}, {Vernieri}, {Walker},
  {Weinberg}, {Young}, {Anthony-Brumfield}, {Bizyaev}, {Breslauer}, {De Lee},
  {Downey}, {Halverson}, {Huehnerhoff}, {Klaene}, {Leon}, {Long}, {Mahadevan},
  {Malanushenko}, {Nguyen}, {Owen}, {S{\'a}nchez-Gallego}, {Sayres}, {Shane},
  {Shectman}, {Shetrone}, {Skinner}, {Stauffer}, \& {Zhao}}]{wilson2019}
{Wilson} J.~C. {et~al.}, 2019, \pasp, 131, 055001

\bibitem[{{Zasowski} {et~al}\mbox{.}(2017){Zasowski}, {Cohen}, {Chojnowski},
  {Santana}, {Oelkers}, {Andrews}, {Beaton}, {Bender}, {Bird}, {Bovy},
  {Carlberg}, {Covey}, {Cunha}, {Dell'Agli}, {Fleming}, {Frinchaboy},
  {Garc{\'{\i}}a-Hern{\'a}ndez}, {Harding}, {Holtzman}, {Johnson}, {Kollmeier},
  {Majewski}, {M{\'e}sz{\'a}ros}, {Munn}, {Mu{\~n}oz}, {Ness}, {Nidever},
  {Poleski}, {Rom{\'a}n-Z{\'u}{\~n}iga}, {Shetrone}, {Simon}, {Smith},
  {Sobeck}, {Stringfellow}, {Szigeti{\'a}ros}, {Tayar}, \&
  {Troup}}]{zasowski2017}
{Zasowski} G. {et~al.}, 2017, \aj, 154, 198

\bibitem[{{Zasowski} {et~al}\mbox{.}(2013){Zasowski}, {Johnson}, {Frinchaboy},
  {Majewski}, {Nidever}, {Rocha Pinto}, {Girardi}, {Andrews}, {Chojnowski},
  {Cudworth}, {Jackson}, {Munn}, {Skrutskie}, {Beaton}, {Blake}, {Covey},
  {Deshpande}, {Epstein}, {Fabbian}, {Fleming}, {Garcia Hernandez}, {Herrero},
  {Mahadevan}, {M{\'e}sz{\'a}ros}, {Schultheis}, {Sellgren}, {Terrien}, {van
  Saders}, {Allende Prieto}, {Bizyaev}, {Burton}, {Cunha}, {da Costa},
  {Hasselquist}, {Hearty}, {Holtzman}, {Garc{\'{\i}}a P{\'e}rez}, {Maia},
  {O'Connell}, {O'Donnell}, {Pinsonneault}, {Santiago}, {Schiavon}, {Shetrone},
  {Smith}, \& {Wilson}}]{zasowski2013}
---, 2013, \aj, 146, 81

\end{thebibliography}
\label{lastpage}
\end{document}